\documentclass[prd,preprintnumbers,superscriptaddress,tightenlines,nofootinbib, eqsecnum]{revtex4-2}

\usepackage{amsmath}
\usepackage{amsfonts}
\usepackage{amssymb}
\usepackage{bm}
\usepackage[colorlinks]{hyperref}
\usepackage{mathrsfs}
\usepackage{graphicx}
\usepackage{empheq}
\usepackage[normalem]{ulem}
\usepackage{tensor}

\usepackage[usenames]{color}
\definecolor{darkgreen}{rgb}{0,0.5,0}
\hypersetup{urlcolor=darkgreen}
\usepackage[capitalize]{cleveref}

\usepackage{ytableau}
\ytableausetup{centertableaux,boxsize=1.6em}

\allowdisplaybreaks

\DeclareSymbolFontAlphabet{\mathrsfs}{rsfs}
\DeclareMathAlphabet{\mathcal}{OMS}{cmsy}{m}{n}

\newcommand{\nm}{\nonumber}
\newcommand{\dd}{\mathrm{d}}
\newcommand{\mpl}{m_\mathrm{Pl}}
\newcommand{\Ldlj}{\Lambda^{(d)}_{\ell,j}}
\newcommand{\alj}{\alpha_{\ell,j}}
\newcommand{\blj}{\beta_{\ell,j}}
\newcommand{\glj}{\gamma_{\ell,j}}
\newcommand{\almj}[1]{\alpha_{\ell-#1,j}}
\newcommand{\blmj}[1]{\beta_{\ell-#1,j}}
\newcommand{\glmj}[1]{\gamma_{\ell-#1,j}}

\begin{document}
 
\title{Multipole expansion at the level of the action in $d$-dimensions}

\author{Loris \textsc{Amalberti}}\email{loris.amalberti@desy.de}
\affiliation{Deutsches Elektronen-Synchrotron DESY, Notkestr. 85, 22607 Hamburg, Germany}
\affiliation{Institut für Physik, Humboldt-Universität zu Berlin, 12489 Berlin, Germany}

\author{Fran\c{c}ois \textsc{Larrouturou}}\email{francois.larrouturou@desy.de}
\affiliation{Deutsches Elektronen-Synchrotron DESY, Notkestr. 85, 22607 Hamburg, Germany}

\author{Zixin \textsc{Yang}}\email{zixin.yang@desy.de}
\affiliation{Deutsches Elektronen-Synchrotron DESY, Notkestr. 85, 22607 Hamburg, Germany}

\date{\today}

\begin{abstract}
In this paper we study the multipole expansion of the long-wavelength effective action for radiative sources in ($d$+1) spacetime dimensions. We present detailed expressions for the multipole moments for the case of scalar-, electromagnetic-, and (linearized) gravitational-wave emission.  For electromagnetism and gravity, we derive expressions for the odd-parity, magnetic-type moments as SO($d$) duals of the ones traditionally used in the literature. The $d$-dimensional case features a novel set of `Weyl-type' moments, coupling to the spatial part of the Weyl tensor, which are absent in three dimensions. Agreement is found in the overlap with previous known results, notably in the $d \to 3$ limit. Due to its reliance on dimensional regularization, the results presented here play a crucial role for the further development of the Effective Field Theory approach to gravitational dynamics, and in particular for the computation of the gravitational-wave flux, starting at the third post-Newtonian order.
\end{abstract}

\preprint{DESY-23-200}

\maketitle

\section{Introduction}\label{sec:intro}

Highly accurate analytic predictions are of prime importance for the signal analysis for gravitational-wave (GW) detectors, notably when it comes to observing the inspiral phase of binary compact objects.
If the current ground-based LIGO-Virgo-KAGRA network is mainly sensitive to rapidly coalescing black holes binaries~\cite{LIGOScientific:2021djp}, this will not be the case for future generations of detectors.
Indeed, both the spaceborne LISA instrument~\cite{amaroseoane2017laser} and the ground-based Einstein Telescope~\cite{Punturo:2010zz, Branchesi:2023mws} are expected to be quite sensitive to the inspiral phase (see~\cite{LISAConsortiumWaveformWorkingGroup:2023arg} in the case of LISA).
It is thus crucial to provide accurate analytic waveforms for the data analysis of those detectors.\vskip 4pt
When it comes to precise analytic predictions for the two-body gravitational problem, the post-Newtonian (PN) approach is a paramount tool.
Focusing on the weak-field and low-velocity inspiral phase of merging compact objects, it allows us to derive the phase evolution and GW amplitude perturbatively to the desired order in $v/c$ (the relative velocity over the speed of light). We let the reader refer to~\cite{Blanchet:2013haa,Buonanno:2014aza,Porto:2016pyg,Goldberger:2022ebt} for reviews on the topic. For non-spinning bodies, the current state-of-the art is the 4.5PN precision for the phase~\cite{Blanchet:2023bwj} (\emph{i.e.} the $(v/c)^9$ correction to the leading order), the 4PN precision for both the GW flux and the dominant quadrupolar amplitude mode~\cite{Blanchet:2023sbv} and the 3.5PN precision for the sub-leading ones~\cite{Faye:2012we,Henry:2021cek,Henry:2022ccf}. For the case of spinning bodies, on the other hand, the state of the art is at 4PN for the GW flux \cite{Cho:2021mqw,Cho:2022syn} and to 3.5PN order for the amplitude \cite{Porto:2012as,Henry:2022dzx,Henry:2023tka}. These results were derived through a combination of techniques, including the Post-Newtonian Multipolar-Post-Minkowkian (PN-MPM) framework~\cite{Blanchet:1985sp,Blanchet:1986dk,Blanchet:1987wq,Blanchet:1992br,Blanchet:1998in} (notably for the non-spinning case), which relies on a careful matching between a PN expansion in the vicinity of the source and a MPM one outside the source; and the effective field theory (EFT) approach~\cite{Goldberger:2004jt,Porto:2005ac,Goldberger:2009qd,Porto:2010zg,Ross:2012fc}, which also relies on a multipolar expansion, together with a systematic separation of the relevant scales of the problem, but directly at the level of the (effective) action \cite{Porto:2016pyg, Goldberger:2022ebt}.  Although the EFT approach has also achieved the 4PN order of accuracy, or next-to-next-to-next-to-next-to-leading order (NNNNLO), in the {\it conservative} sector for non-spinning bodies \cite{Galley:2015kus,Foffa:2019yfl,Foffa:2019rdf} (see also \cite{Blanchet:2019rjs,Almeida:2021xwn,Almeida:2023yia,Henry:2023sdy,Blumlein:2020pyo,Levi:2022rrq} for results at higher orders), the computation of the GW flux has been performed only to NNLO, at 2PN~\cite{Leibovich:2019cxo}. In~order to move forward, towards higher levels of accuracy, the well-known divergences that appear already at 3PN, both in the equations of motion~\cite{Foffa:2011ub} and in the non-linear radiative corrections~\cite{Goldberger:2009qd}, must be carefully tackled. 
Within dimensional regularization, extensively used in the EFT approach since the seminal work of \cite{Goldberger:2004jt} (see also~\cite{Blanchet:2003gy,Blanchet:2005tk}), divergences arise as poles $\propto (d-3)^{-1}$, with $d$ the number of spatial dimensions. Even though these divergences can be carefully removed from observable quantities in the conservative sector to 4PN order~\cite{Foffa:2019rdf}, the computation of the GW flux requires a careful analysis of the multipole expansion in $d$-dimensions.\vskip 4pt The multipole expansion at the level of the action in three dimensions was originally performed in~\cite{Ross:2012fc}. The purpose of this paper is to extend those results to the case of an arbitrary number of spatial dimensions. Building upon the analysis in~\cite{Ross:2012fc} we study the scalar, electromagnetic and (linearized) gravitational cases, in that order. Along the way, we also verify that the three-dimensional limits of our results are consistent with those exposed in~\cite{Damour:1990gj, Ross:2012fc}. As it was argued in~\cite{Henry:2021cek}, an important subtlety arises when considering odd-parity (\emph{i.e.} magnetic-type) moments. In three spatial dimensions, those are constructed as irreducible representations of SO(3) \emph{via} contractions between purely symmetric and trace-free (STF) tensors and a Levi-Civita symbol~\cite{Blanchet:1998in,Ross:2012fc}.
Such feature, however, is specific to $d=3$, since there is no simple generalization of the Levi-Civita symbol to arbitrary dimensions.
Therefore, when deriving multipole moments as irreducible representations of SO($d$), we must consider all possible Young tableaux, and magnetic moments will carry non-trivial symmetry properties described by a mixed Young tableaux~\cite{james1987representation,ma2004problems,Bekaert:2006py}.
Additionally, a new set of multipole moments emerges, corresponding to a different mixed Young tableaux, which does not exist in three dimensions.
We point the interested reader to~\cite{Henry:2021cek} for a more detailed discussion about this subtle point, and to~\cite{Almeida:2021xwn,Almeida:2023yia,Henry:2023sdy} for some applications in the conservative sector. The calculation of the GW flux to 3PN order within the EFT approach, where the results derived here are of utmost relevance, will be reported elsewhere.\vskip 4pt
This work is organized as follows. The $d$-dimensional multipolar expansion of a scalar field is presented in Sec.~\ref{sec:scalar}, the electromagnetic case is treated in Sec.~\ref{sec:electromag}, and gravity, in Sec.~\ref{sec:gravity}. Sec.~\ref{sec:concl} concludes this work.
Useful decomposition formulas are collected in App.~\ref{sec:app_irred_decomp} and identities coming from conservation laws, in App.~\ref{sec:app_conservation}. Finally, cumbersome computations that are too long to be presented in the main text are displayed in App.~\ref{sec:app_technical_details}.\vskip 4pt
\textbf{Notation:} We use natural units $c=1=\hbar$, and work in a spacetime with one time and $d$ spatial dimensions, equipped with a mostly negative metric signature. Greek letters denote Lorentz indices (running from 0 to $d$) and Latin letters, spatial ones (running from 1 to $d$).
Bold symbols denote spatial vectors, \emph{e.g.} $\mathbf{x} = x^i$, and we define the d'Alembertian operator on the flat, Minkowskian background, as $\Box \equiv \eta^{\mu\nu}\partial_\mu\partial_\nu= \partial_t^2 - \nabla^2$.
We employ the multi-index notation as introduced in~\cite{Blanchet:1985sp}, \emph{i.e.} $x^L \equiv x^{i_1}x^{i_2}\dots x^{i_{\ell-1}}x^{i_{\ell}}$ and $I^L  \equiv I^{i_1i_2 \dots i_{\ell-1}i_\ell}$, and weight the (anti-)symmetrizations, \emph{e.g.} $T^{(L)} = \underset{L}{\mathcal{S}}\left(T^L\right) = \frac{1}{\ell !}\left(T^L+\ell\text{-permutations}\right)$, or $T^{[ab]} = \underset{ab}{\mathcal{A}}\left(T^{ab}\right) = \frac{1}{2}\left(T^{ab} - T^{ba}\right)$. The symmetric trace-free (STF) operator is denoted with hats or brackets, as $\hat{T}^L = T^{\langle L \rangle} \equiv \underset{L}{\text{STF}} \left(T^L\right)$.
Last but not least, we follow the notation in \cite{Henry:2021cek} for the magnetic- and Weyl-like multipole moments, introduced in Sec.~\ref{sec:electromag} and~\ref{sec:gravity}, that correspond to the mixed Young tableaux.

\section{Scalar Field}\label{sec:scalar}

Let us start by investigating the simplest case of a scalar field $\phi$, linearly coupled to a source $J$ in a ($d$+$1$)-dimensional spacetime. The corresponding action reads
\begin{equation}
S_{\phi} = \int\!\!\dd t\int\!\!\dd^d\mathbf{x}\,\left(\frac{1}{2}\,\partial_{\mu}\phi\,\partial^{\mu}\phi+J\phi \right)\,,
\end{equation}
and the equation of motion (EOM) outside the source is given by
\begin{equation}\label{ScalEOM}
\Box \phi=0\,.
\end{equation}
We assume that the source is compact-supported, with typical size $a$, and that the spatial evolution of the field outside the source is described by a typical scale $\lambda$.
Hereafter, we work in the \emph{long wavelength approximation}, \emph{i.e.} in the regime where $a \ll \lambda$ holds.
In this framework, we are allowed to perform a Taylor expansion of the scalar field around a point in space within the source, which for simplicity we choose to coincide with the origin of our coordinate system. 
This translates in
\begin{equation}
\phi\left(t,\mathbf{x}\right)
=\sum_{n=0}^{\infty}\,\frac{1}{n!}\,x^N\left(\partial_N \phi\right)\left(t,\mathbf{0}\right)\,,
\end{equation}
which we then plug into the source term of the action
\begin{equation}
S_\text{source} = \int\!\!\dd t\int\!\!\dd^d\mathbf{x}\,J\left(t,\mathbf{x} \right)\phi\left(t,\mathbf{x} \right)  =
\int\!\!\dd t\,\sum_{n=0}^{\infty}\,\frac{1}{n!}\left(\int\!\!\dd^d\mathbf{x}\,J\left(t,\mathbf{x} \right)x^N\right)\partial_N\phi\,,
\label{ScalExpanded}
\end{equation}
where $\partial_N\phi = \left(\partial_N \phi\right)\left(t,\mathbf{0}\right)$.
We can already recognize a multipolar structure, where the multipole moments are given by the coefficients of the $\partial_N \phi$ operators. 
We now need to express those moments as irreducible representations of the rotation group SO($d$). The formula for an arbitrary symmetric tensor $S^N$ expressed in terms of fully STF tensors is given by \cite{thorne1980multipole, Blanchet:2003gy}
\begin{align}
	S^{N} = \sum^{\left[n/2\right]}_{p=0} \,\frac{n!}{\left(n-2p\right)!}\,\Lambda^{(d)}_{n-2p,p}\,\delta^{(i_{1}i_{2}}...\,\delta^{i_{2p-1}i_{2p}}\,\hat{S}^{i_{2p+1}...i_{n})a_{1}a_{1}...a_{p}a_{p}}\,,
\label{eq:STFformula}
\end{align}
where $[n/2]$ denotes the integer part of $n/2$ and we defined the coefficients
\begin{equation}
\Lambda^{(d)}_{n,p} \equiv \frac{\Gamma \left( \frac{d}{2}+n\right)}{2^{2p}p!\,\Gamma \left( \frac{d}{2}+n+p\right)}\,
\label{lambda}\,.
\end{equation}
In particular, we express the fully symmetric structures $x^N$ in terms of their STF counterparts
\begin{equation}
x^N=\sum_{p=0}^{\left[n/2\right]}\,\frac{n!}{\left(n-2p\right)!}\,\Lambda^{(d)}_{n-2p,p}\,\delta^{(i_{1}i_{2}}...\delta^{i_{2p-1}i_{2p}}\hat{x}^{i_{2p+1}...i_{n})}r^{2p}\,,
\end{equation}
with $r=|\mathbf{x}|$ and we substitute into~\eqref{ScalExpanded}, which now reads
\begin{align}
S_{\text{source}} 
& \nm
=\int\!\!\dd t\,\sum_{n=0}^{\infty}\sum^{\left[n/2\right]}_{p=0}\,\frac{\Lambda^{(d)}_{n-2p,p}}{\left(n-2p\right)!}\,\int\!\!\dd^d\mathbf{x} J \hat{x}^{N-2P} r^{2p}\hat{\partial}_{N-2P} \,\nabla^{2p}\phi\\
&
=\int\!\!\dd t\,\sum_{\ell,\,j=0}^{\infty}\,\frac{\Ldlj}{\ell!}\,\int\!\!\dd^d\mathbf{x}\,\partial_t^{2j} J \hat{x}^{L} r^{2j}\hat{\partial}_L \phi\,,
\label{eq:nablaeom}
\end{align}
where we used ~\eqref{ScalEOM} in \eqref{eq:nablaeom} to exchange the Laplacian operators into time derivatives on the fields, which are then shifted onto the source moment $J$ via integration-by-parts.
It is now trivial to read off the sought structure
\begin{equation}
S_{\text{source}}=\int\!\!\dd t\,\sum_{\ell=0}^{\infty}\, \frac{1}{\ell !}\, I^L\partial_L\phi\,,
\end{equation}
with multipole moments given by irreducible representations of SO($d$) as
\begin{equation}
I^L= \sum_{j=0}^{\infty}\,\frac{\Gamma \left( \frac{d}{2}+\ell\right)}{2^{2j}j!\,\Gamma \left( \frac{d}{2}+\ell+j\right)}\,\int\!\!\dd^d\mathbf{x}\,\partial_t^{2j} J r^{2j} \hat{x}^{L}\,.\label{ScalMulti}
\end{equation}
In three dimensions, the combination $\Ldlj$ becomes
\begin{equation}
\Lambda^{(d=3)}_{\ell,j}=\frac{\left(2\ell+1 \right)!!}{\left(2j \right)!!\left(2\ell+2j+1\right)!!}\,,\label{lambda3d}
\end{equation}
hence the three-dimensional limit of our result, ~\eqref{ScalMulti}, is fully consistent with the known three-dimensional multipole expansion of a scalar field, \emph{e.g.} Eq.~(10) of~\cite{Ross:2012fc}.

\section{Electromagnetism}\label{sec:electromag}

\subsection{Framework description}\label{sec:EM_framework}

An electromagnetic field $A_{\mu}$ linearly coupled to a source $J^{\mu}$ in a ($d$+$1$)-dimensional spacetime is described by the following action,
\begin{equation}
S_{\text{EM}}= -\int\!\!\dd t\int\!\!\dd^d\mathbf{x}\,\left(\frac{1}{4}\,F^{\mu \nu}F_{\mu \nu}+J^{\mu}A_{\mu} \right)\,,
\label{eq:S_EM}
\end{equation}
where $F_{\mu \nu}\equiv \partial_{\mu}A_{\nu}-\partial_{\nu}A_{\mu}$ is the usual field strength tensor. The current $J^\mu$ is conserved, i.e. $\partial_{\alpha}J^{\alpha}=0$. The field strength can be further decomposed in terms of the electric and magnetic fields, 
\begin{equation}
E_{a}\equiv F_{a0} =\partial_{a}A_{0}-\partial_tA_{a}\,,
\qquad
B_{a|b}\equiv F_{ab}=\partial_{a}A_{b}-\partial_{b}A_{a}\,.
\end{equation}

Instead of the usual magnetic field in three dimensions $B_a = \varepsilon_{abc}F_{bc}/2$, we adopt its dual $B_{a\vert b}$ to avoid the ambiguity of Levi-Civita symbols in generic dimensions.
In vacuum space where $J^\mu = 0$, the equations of motion, Maxwell equations and Bianchi identity for the electromagnetic field are given by 
\begin{equation}\label{MaxDD}
\Box F_{\mu \nu}=0\,,
\qquad
\partial_{\alpha}F^{\alpha \beta}=0
\qquad\text{and}\qquad
\partial_{[\alpha}F_{\beta \sigma]}=0\, ,
\end{equation}
respectively, which can also be written as
\begin{equation}\label{MaxSpatial}
\partial_a E_a=0\,,\qquad \partial_a B_{a|b}=\partial_t E_{b}\,,\qquad 2\,\partial_{[a}E_{b]}=\partial_t B_{a|b}\,,\qquad \Box E_a= \Box B_{a|b}=0\,,
\end{equation}
in terms of the electric and magnetic fields.

\subsection{Split of the action}\label{sec:EM_split}

Assuming a compact-supported source, we work in the long wavelength approximation. The electromagnetic field can be safely Taylor-expanded as
\begin{equation}
A^{\mu}\left(t,\mathbf{x}\right)=\sum_{n=0}^{\infty}\,\frac{1}{n!}\,x^N\left(\partial_N A^{\mu}\right)\left(t,\mathbf{0}\right)\,.
\end{equation}
Plugging the Taylor expansion of the field into \eqref{eq:S_EM}, the source term of the action becomes
\begin{align}\label{emExpanded}
S_{\text{source}}
= & \nm
-\int\!\!\dd t\,\int\!\!\dd^d\mathbf{x}\,J^{\mu}\left(t,\mathbf{x} \right)\sum_{n=1}^{\infty}\,\frac{1}{n!}\,x^N\partial_N A_{\mu}\\
= & \nm
-\int\!\!\dd t\left(\int\!\!\dd^d\mathbf{x}\,J^{0}\right)A_{0}\\
& 
-\int\!\!\dd t\,\sum_{n=1}^{\infty}\,\frac{1}{n!}\left(\,\int\!\!\dd^d\mathbf{x}\,J^0x^N\right)\partial_N A_0
-\int\!\!\dd t\,\sum_{n=0}^{\infty}\,\frac{1}{n!}\left(\,\int\!\!\dd^d\mathbf{x}\,J^bx^N\right)\partial_{N}A_b\,.
\end{align}
In the second equality the expansion of $A_0$ is separated into two sectors.
The first term which is free of derivatives, is nothing but the monopole representing the coupling of the field to the total electric charge $Q \equiv \int\!\!\dd^d\mathbf{x}\,J^{0}$.
This term does not radiate, and thus is singled out from the multipole expansion.
The last two terms encrypt radiative modes, which should couple to the two propagating degrees of freedom, $E_a$ and $B_{a\vert b}$ collectively. For this purpose, the last coefficient in the action~\eqref{emExpanded} can be conveniently expressed in terms of its corresponding irreducible decomposition utilizing Young symmetrizers~\cite{james1987representation,ma2004problems,Bekaert:2006py}, here denoted as Young tableaux through a slight abuse of notation
\begin{align}
\int\!\!\dd^d\mathbf{x}\,J^bx^N &=
\frac{1}{\left(n+1\right)!}\,
\ytableausetup{mathmode, boxframe=normal, boxsize=1.67em}
\begin{ytableau}[]
\text{\footnotesize{$b$}} & \text{\footnotesize{$i_1$}}  & \text{\footnotesize{$\ldots$}} & \text{\footnotesize{$i_n$}} 
\end{ytableau}
+
\frac{n}{\left(n+1\right)!}
\,\left(\,\,\begin{ytableau}[]
\text{\footnotesize{$b$}} & \text{\footnotesize{$i_1$}}  & \text{\footnotesize{$\ldots$}} &  \text{\footnotesize{$i_{n-1}$}} \\
i_n & \none & \none & \none
\end{ytableau}+\text{$i$-perms}\right)\nonumber\\
&=
\int\!\!\dd^d\mathbf{x}\,J^{(b}x^{N)}+\frac{2 n}{n+1}\,\underset{N}{\mathcal{S}}\left(\int\!\!\dd^d\mathbf{x}\,J^{[b}x^{i_n]N-1}\right)\,,
\end{align}
where, in the first equality, ``+$i$-perms'' means that all index combinations $\{i_1, \ldots, i_n\}$ must be added together.
Implementing this decomposition and using the conservation law~\eqref{Jcons2}, the last term of the action~\eqref{emExpanded} can then be rewritten as
\begin{align}
S^{A_b}_{\text{source}}
= & \nm 
-\int\!\!\dd t\,\sum_{n=0}^{\infty}\,\frac{1}{n!}\left(\,\int\!\!\dd^d\mathbf{x}\,J^{(b}x^{N)}\right)\partial_{N}A_b-\int\!\!\dd t\,\sum_{n=1}^{\infty}\,\frac{2n}{\left(n+1\right)!}\left(\,\int\!\!\dd^d\mathbf{x}\,J^{[b}x^{i_n]N-1}\right)\partial_{N}A_b\\
= & 
\int\!\!\dd t \,\sum_{n=1}^{\infty}\,\frac{1}{n!}\left(\,\int\!\!\dd^d\mathbf{x}\, J^{0}x^{N}\right)\partial_{N-1}\partial_t A_{i_n}+\int\!\!\dd t\,\sum_{n=1}^{\infty}\,\frac{n}{\left(n+1\right)!}\left(\,\int\!\!\dd^d\mathbf{x}\,J^{b}x^{N}\right)\partial_{N-1}B_{b|i_n}\,.
\end{align}
With this expression at hand, the source action is now split as
\begin{equation}
S_{\text{source}}=S^{\text{cons}}_{\text{source}}+S^{\text{rad}}_{\text{source}}\,,
\end{equation}
with
\begin{subequations}\label{eq:SEM_source}
\begin{align}
S^{\text{cons}}_{\text{source}} =
&
-\int\!\!\dd t \,Q \,A_{0}\,,\\
\label{Jsource}
S^{\text{rad}}_{\text{source}} =
&\nm
\int\!\!\dd t\,\sum_{n=1}^{\infty}\,\frac{1}{n!}\left(\,\int\!\!\dd^d\mathbf{x}\,J^0x^N\right)\partial_{N-1} E^{i_n}\\
&
+\int\!\!\dd t\,\sum_{n=1}^{\infty}\,\frac{n}{\left(n+1\right)!}\left(\,\int\!\!\dd^d\mathbf{x}\,J^{a}x^{N}\right)\partial_{N-1}B_{a|i_n}\,.
\end{align}
\end{subequations}
Just as in the scalar field case, a multipolar structure starts to manifest, which is yet to be expressed in terms of irreducible representations of SO($d$).\\

Before moving on to such reduction in the next section, let us point out the consistency of the three-dimensional limit of the expansion~\eqref{eq:SEM_source} with known results.
The monopole term as well as the $J^0$ sector are trivial. 
As for the $J^a$ sector, 
in three dimensions any SO($3$) antisymmetric rank-$2$ tensor can be traded for its dual vector counterpart (see \emph{e.g.}~\cite{james1987representation}).
Hence, we can define the three-dimensional magnetic field $B_a$ as the limit of the dual of $B_{a\vert b}$, by
\begin{equation}
\lim_{d \to 3}  B_{a|b} \equiv  \varepsilon_{abc}B_c \quad  \Leftrightarrow \quad
B_a \equiv  \frac{1}{2}\varepsilon_{abc} \,\lim_{d \to 3}B_{b|c}\,,
\label{mag3d}
\end{equation}
where $\varepsilon_{abc}$ is the three-dimensional Levi-Civita symbol. 
By injecting this limit in the last line of ~\eqref{Jsource}, in three dimensions the magnetic sector reduces to
\begin{align}
\lim_{d\to 3} S^{B_{a\vert i_n}}_\text{source}
= & \nm
\lim_{d\to 3} \int\!\!\dd t\,\sum_{n=1}^{\infty}\,\frac{n}{\left(n+1\right)!}\left(\,\int\!\!\dd^d\mathbf{x}\,J^{a}x^{N}\right)\partial_{N-1}B_{a|i_n}\\
= & \nm
\int\!\!\dd t\,\sum_{n=1}^{\infty}\,\frac{n}{\left(n+1\right)!}\left(\,\int\!\!\dd^3\mathbf{x}\left(J^{a}x^{b}\right)x^{N-1}\right)\partial_{N-1}\left(\varepsilon_{a b c}B_c\right)\\
= & 
\int\!\!\dd t\,\sum_{n=1}^{\infty}\,\frac{n}{\left(n+1\right)!}\left(\,\int\!\!\dd^3\mathbf{x}\left(\mathbf{J}\times \mathbf{x}\right)^ax^{N-1}\right)\partial_{N-1}B_{a}\,.
\end{align}
Such expression is the usual form of the magnetic expansion in three dimensions, see \emph{e.g.} Eq.~(35) of~\cite{Ross:2012fc}. 

\subsection{Irreducible decomposition of the moments}\label{sec:EM_decomp}
Let us now express the moments appearing in ~\eqref{Jsource} in terms of irreducible representations of SO($d$).
As they are of different nature, we treat the scalar sector composed of the $J^0$ term, and the vector one involving $J^a$ separately.

Consider the scalar sector and apply the relations~\eqref{xSTF2} and~\eqref{eq:xLdL}, which leads to
\begin{align}\label{SJ0}
S_{\,\text{rad}}^{J^0}
= & \nm
\int\!\!\dd t\,\sum_{n=1}^{\infty}\,\frac{1}{n!}\left(\,\int\!\!\dd^d\mathbf{x}\,J^0x^N\right)\partial_{N-1} E^{i_n}\\
= & \nm
\int\!\!\dd t\,\sum_{\ell,\,j=0}^{\infty}\,\frac{\Ldlj}{\ell!\,\left(\ell+2j+1\right)}\,\int\!\!\dd^d\mathbf{x}\,\partial_t^{2j}J^0x^a\hat{x}^Lr^{2j}\,\hat{\partial}_L E^a\\
= & 
\int\!\!\dd t\,\sum_{\ell=1}^{\infty}\,\sum_{j=0}^{\infty}\,\frac{\Ldlj}{\left(\ell-1\right)!\,\left(\ell+2j\right)}\,\int\!\!\dd^d\mathbf{x}\,\partial_t^{2j}J^0\hat{x}^{aL-1}r^{2j}\,\hat{\partial}_{L-1} E^a\,,
\end{align}
which is already in the desired STF form.
Next, we move on to the moments involving $J^a$, 
\begin{equation}
S_{\text{rad}}^{J^a} 
=
-\int\!\!\dd t\,\sum_{n=1}^{\infty}\,\frac{n}{\left(n+1\right)!}\left(\,\int\!\!\dd^d\mathbf{x}\,J^ax^{N}\right)\partial_{N-1}B_{a|i_n}\,.
\end{equation}
The first step is to express the purely symmetric structure $x^N$ in terms of its STF counterpart, $\hat{x}^N$. Use the STF relations~\eqref{xSTF1} and~\eqref{eq:xLdL} we obtain
\begin{align}
\label{JbFIRST}
 S_{\text{rad}}^{J^a} 
= & \nm
\int\!\!\dd t\,\sum_{\ell,j=0}^{\infty}\,\frac{\Ldlj}{\ell!\,\left(\ell+2j+2\right)}\,\int\!\!\dd^d\mathbf{x}\,\partial_t^{2j}J^ax^b\hat{x}^Lr^{2j}\,\hat{\partial}_L B_{a|b}\\
= & \nm
\int\!\!\dd t\,\sum_{\ell=1}^{\infty}\,\sum_{j=0}^{\infty}\,\frac{\Ldlj}{\left(\ell-1\right)!\,\left(\ell+2j+1\right)}\,\int\!\!\dd^d\mathbf{x}\,\partial_t^{2j}J^a\hat{x}^{bL-1}r^{2j}\,\hat{\partial}_{L-1} B_{a|b}\\
+& 
\int\!\!\dd t\,\sum_{\ell=1}^{\infty}\,\sum_{j=0}^{\infty}\,\frac{\Ldlj}{\left(\ell-1\right)!\,\left(\ell+2j+2\right)\left(d+2\ell-2\right)}\,\int\!\!\dd^d\mathbf{x}\,\partial_t^{2j+1}J^a\hat{x}^{L-1}r^{2j}\,\hat{\partial}_{L-1} E^a\,.
\end{align}
The action is not yet in the irreducible form at this stage, due to the vectorial nature of $J^a$. 
We hence need to reduce it more towards fully irreducible representations of SO($d$).
After some cumbersome derivation presented in details in App.~\ref{sec:app_EMSTF}, the final result is given by
\begin{align}\label{sourcemultexpEM}
S_{\text{source}}
= &\nm
S_{\text{cons}}+S_{\text{rad}}^{J^0}+S_{\text{rad}}^{J^a}\\
= & 
-\int\!\!\dd t \,Q \,A_{0}+\int\!\!\dd t \,\sum_{\ell=1}^{\infty}\,\frac{1}{\ell !}\,I^L\,\partial_{L-1}E^{i_{\ell}}+\int\!\!\dd t \,\sum_{\ell=1}^{\infty}\,\frac{\ell}{\left(\ell+1\right) !}\,J^{a|L}\,\partial_{L-1}B_{a|i_{\ell}}\,,
\end{align}
with the $d$-dimensional electric and magnetic multipole moments reading respectively
\begin{subequations}
\label{EMMOM}
\begin{align}
\label{EMelectricMOM}
I^L
= & \nm
\sum_{j=0}^{\infty}\,\frac{\Gamma \left( \frac{d}{2}+\ell\right)}{2^{2j}j!\,\Gamma \left( \frac{d}{2}+\ell+j\right)}\left(1+\frac{2j}{d+\ell-2}\right)\int\!\!\dd^d\mathbf{x}\,\partial_t^{2j}J^0\hat{x}^{L}r^{2j}\\
& 
-\sum_{j=0}^{\infty}\,\frac{\Gamma \left( \frac{d}{2}+\ell\right)}{2^{2j}j!\,\Gamma \left( \frac{d}{2}+\ell+j\right)}\frac{1}{\left(d+\ell-2\right)}\,\int\!\!\dd^d\mathbf{x}\,\partial_t^{2j+1}\tilde{J}\hat{x}^{L}r^{2j}\,,\\
\label{EMmagneticMOM}
J^{a|L} 
= &\
\underset{a i_{\ell}}{\mathcal{A}}\,\sum_{j=0}^{\infty}\,\frac{\Gamma \left( \frac{d}{2}+\ell\right)}{2^{2j}j!\,\Gamma \left( \frac{d}{2}+\ell+j\right)}\left[\int\!\!\dd^d\mathbf{x}\,\partial_t^{2j}J^{a}\hat{x}^{L}r^{2j}\right]^{\text{TF}}\,,
\end{align}
\end{subequations}
where $\tilde{J} \equiv J^ax^a$.
The electric and magnetic moments are indeed irreducible representations of SO($d$), as their symmetries are respectively given by the symmetric and mixed Young tableaux~\cite{james1987representation,ma2004problems,Bekaert:2006py}
\begin{equation}
I^L =
\ytableausetup{mathmode, boxframe=normal, boxsize=1.7em}
\begin{ytableau}[]
i_\ell & i_{\ell-1}  & \ldots & i_2 & i_1 
\end{ytableau}
\qquad\text{and}\qquad
J^{a\vert L} =
\ytableausetup{mathmode, boxframe=normal, boxsize=1.7em}
\begin{ytableau}[]
i_\ell & i_{\ell-1}  & \ldots & i_2 & i_1 \\
a & \none & \none & \none & \none
\end{ytableau}\,.
\end{equation}

In the $d=3$ limit, ~\eqref{EMMOM} fully agrees with the known three-dimensional multipole expansion results.
It is trivial to recognize that the electric multipole~\eqref{EMelectricMOM} reduces to its three-dimensional counterpart, (47) of~\cite{Ross:2012fc}, whereas comparing magnetic moments requires more work.
In the three-dimensional limit, one can decompose the antisymmetric structure of ~\eqref{EMmagneticMOM} as a product of Levi-Civita symbols, leading to
\begin{align}\label{EMmagneticMOM3d}
\lim_{d\to 3}J^{a|L} = &\nm
 \frac{1}{2}\,\varepsilon^{cai_{\ell}}\sum_{j=0}^{\infty}\,\frac{\left(2\ell+1 \right)!!}{\left(2j \right)!!\left(2\ell+2j+1\right)!!}\left[\int\!\!\dd^d\mathbf{x}\,\partial_t^{2j}\left(\varepsilon^{cpq}J^{p}x^{q}\right)\hat{x}^{L-1}r^{2j}\right]^{\text{TF}}\\
= &\ 
\frac{1}{2}\,\varepsilon^{cai_{\ell}}J_{d=3}^{cL-1}\,,
\end{align}
where we recover the three-dimensional expression of the magnetic moment, Eq.~(48) of~\cite{Ross:2012fc}
\begin{equation}
J_{d=3}^{L}=\sum_{j=0}^{\infty}\,\frac{\left(2\ell+1 \right)!!}{\left(2j \right)!!\left(2\ell+2j+1\right)!!}\,\int\!\!\dd^d\mathbf{x}\,\partial_t^{2j}\left(\mathbf{J}\times\mathbf{x}\right)^{\langle i_{\ell}}\hat{x}^{L-1\rangle}r^{2j}\,.
\end{equation}
Hence, the magnetic sector of the action reduces to
\begin{equation}
\lim_{d\to 3}S^{\text{magnetic}}_\text{rad} =
\lim_{d\to 3}\int\!\!\dd t \,\sum_{\ell=1}^{\infty}\,\frac{\ell}{\left(\ell+1\right) !}\,J^{a|L}\,\partial_{L-1}B_{a|i_{\ell}} = \int\!\!\dd t \,\sum_{\ell=1}^{\infty}\,\frac{\ell}{\left(\ell+1\right) !}\,J^L_{d=3}\,\partial_{L-1}B_{i_{\ell}}\,,
\end{equation}
where we recall that the three-dimensional magnetic field $B_a$ is defined in ~\eqref{mag3d}.
This limit is in full agreement with the known three-dimensional result.

\section{Linearized Gravity}\label{sec:gravity}

\subsection{Framework description}\label{sec:Grav_framework}

Let us now consider the linearized approximation to General Relativity, by perturbing the metric around a flat background as
\begin{equation}
g_{\alpha \beta}=\eta_{\alpha \beta}+\frac{h_{\alpha \beta}}{\mpl}\,,
\end{equation}
where $\eta_{\alpha \beta}$ is the Minkowski metric and the reduced Planck mass reads $\mpl^2=1/\left(32 \pi G\right)$.
From the usual Christoffel symbols $\Gamma^\mu_{\nu\rho}$, we define the Riemann tensor as
\begin{equation}
R^{\mu}_{\ \nu \alpha \beta}=\partial_{\alpha}\Gamma^{\mu}_{\nu \beta}-\partial_{\beta}\Gamma^{\mu}_{\nu \alpha}+\Gamma^{\mu}_{\alpha \tau} \Gamma^{\tau}_{\nu \beta}-\Gamma^{\mu}_{\beta \tau} \Gamma^{\tau}_{\nu \alpha}\,,
\end{equation}
from which the Ricci tensor $R_{\alpha \beta}=R^{\mu}_{\ \alpha \mu \beta}$ and Ricci scalar $R=g^{\alpha \beta}R_{\alpha \beta}$ are obtained.

We restrain ourselves to the linear approximation, implementing a coupling between the graviton and a compact supported source, as
\begin{equation}
S=
-2 \mpl^2\,\int\!\!\dd t \int\!\! \dd^d \mathbf{x}\, \sqrt{-g}\,R-\frac{1}{2 \mpl}\,\int\!\!\dd t \int\!\! \dd^d \mathbf{x}\, T^{\mu \nu}h_{\mu \nu}\,,
\end{equation}
comprised of the Einstein-Hilbert action and a linearized source term. The source term is conserved at linear level, $\partial_\alpha T^{\alpha \beta} = \mathcal{O}(h)$.
The vacuum equations of motion leads to
\begin{equation}\label{GRMaxDD}
R_{\alpha \beta}=0\,,\qquad
\partial_{\alpha}R^{\alpha}_{\ \beta \mu \nu}=0\,,
\qquad
\partial_{[\sigma}R_{\alpha \beta] \mu \nu}=0
\qquad\text{and}\qquad
\Box R_{\alpha \beta \mu \nu}=0\,.
\end{equation}
The Riemann tensor can be further split into propagating degrees of freedom, depending on their parity under SO($d$), as
\begin{subequations}\label{expEBW}
\begin{align}
&
E_{ab}\equiv R_{0a0b}=\frac{1}{2 \mpl}\bigg( \partial_a\partial_t h_{0b}+\partial_b\partial_th_{0a}-\partial^2_th_{ab}-\partial_a\partial_bh_{00}\bigg)\,,\\
&
B_{a|bc}\equiv R_{bac0}=\frac{1}{2 \mpl}\bigg( \partial_a\partial_c h_{0b}+\partial_b\partial_th_{ac}-\partial_b\partial_c h_{0a}-\partial_a\partial_th_{bc}\bigg)\,,\\
&
\mathcal{W}_{abcd}=R_{abcd}+\frac{1}{d-2}\bigg(\delta_{ad}\,E_{bc}+\delta_{bc}\,E_{ad}-\delta_{ac}\,E_{bd}-\delta_{bd}\,E_{ac}\bigg)\,,\label{expWeil}
\end{align}
\end{subequations}
where the Riemann tensor $R_{abcd}$ at the linear order is explicitly given by
\begin{equation}
R_{abcd}=\frac{1}{2 \mpl}\bigg( \partial_b\partial_c h_{ad}+\partial_a\partial_dh_{bc}-\partial_a\partial_ch_{bd}-\partial_b\partial_dh_{ac}\bigg)\,.\label{expRspace3d}
\end{equation}
By analogy with the electromagnetic case, the even-parity $E_{ab}$ and odd-parity $B_{a|bc}$ are respectively dubbed ``electric'' and ``magnetic'' components of the Riemann tensor.
Note that, as advertised previously, we have to deal with the dual of the usual magnetic-type component of the Riemann tensor $B_{a\vert bc}$, which is antisymmetric in $\{a,b\}$ and trace-free in all its indices. Moreover, to avoid confusion, we point out there is no obvious symmetry in $\{b,c\}$. 

In the three-dimensional limit, it reduces to the usual magnetic-type component of the Riemann tensor, $B_{ab}$, as
\begin{equation}\label{grEBlimit}
\lim_{d\to 3} B_{a|bc}=\varepsilon_{abd}B_{cd}
\quad\Leftrightarrow\quad 
B_{ab} = \frac{1}{2}\varepsilon_{cd(a}\lim_{d\to 3}B_{\underline{c}\vert\underline{d} b)}\,,
\end{equation}
where underlined indices are excluded from antisymmetrization.
As for the new component $\mathcal{W}_{abcd}$, it denotes the $d$-dimensional Weyl tensor, and hence bears its particular parity under SO($d$).
Such object should vanish in three dimensions,\footnote{This can be easily understood by considering its SO($3$) dual $C_{ab} \propto \left(\delta_{ab}\,\mathcal{W}^{cd}_{\;\;\,cd}-2\mathcal{W}^c_{\;\;acb} \right)$, which is vanishing as the Weyl tensor is traceless by construction.} as the number of its independent components is given by
\begin{equation}\label{weylComp}
\#\text{ of Weyl components}=\frac{d\left(d+1\right)\left(d+2\right)\left(d-3\right)}{12}\,.
\end{equation}
Hence, in three dimensions the spatial Riemann tensor in terms of $E_{ab}$ Can be expressed as
\begin{equation}
\lim_{d\to 3}R_{abcd} =-\varepsilon_{abe}\varepsilon_{cdf}\,\lim_{d\to 3}E_{ef}\,,
\label{grWeyllimit}
\end{equation}
where the right-hand side involves Levi-Civita symbols. 
Nevertheless, this work takes place in an arbitrary number of spatial dimensions, thus we need to consider $\mathcal{W}_{abcd}$ as being as relevant as $E_{ab}$ or $B_{a\vert bc}$~\cite{Henry:2021cek}.
The three propagating degrees of freedom correspond to the symmetric and mixed Young tableaux as~\cite{james1987representation,ma2004problems,Bekaert:2006py}
\begin{equation}
E_{ab} = 
\begin{ytableau}[]
a & b
\end{ytableau}
\,,
\qquad
B_{a\vert b c} = 
\begin{ytableau}[]
b & c \\
a & \none
\end{ytableau}
\qquad\text{and}\qquad
\mathcal{W}_{abcd} = 
\begin{ytableau}[]
a & c \\
b & d
\end{ytableau}
\,.
\end{equation}
In addition to being obviously traceless from ~\eqref{expEBW}, these propagating degrees of freedom obey Maxwell-like equations, derived from ~\eqref{GRMaxDD}
\begin{subequations}
\label{eq:MaxwellGrav}
\begin{align}
&
E_{aa}=0\,,
& &
B_{a|bb}=0\,,
& &
\mathcal{W}_{abac}=0\,,\label{grTraces}\\
&
\partial_a{E_{ab}}=0\,,
& &
\partial_c{B_{c|ab}}=\partial_t E_{ab}\,,
& &
\partial_c{B_{a|bc}}=0\,,
\label{grMaxlatin1}\\
&
2\,\partial_{[c} E_{a]b}=\partial_t B_{c|ab}\,,
& &
2\,\partial_{[c} B_{\underline{a}|\underline{b}d]} =\partial_t R_{abcd}\,,
& &
\partial_d \mathcal{W}_{cdab}=\frac{\left(d-3 \right)}{\left( d-2\right)}\partial_t B_{b|ac}\,,\label{grMaxlatin2}
\end{align}
\end{subequations}
where underlined indices are again excluded from antisymmetrization.

\subsection{Split of the action}\label{sec:Grav_split}

We assume that the source is compact-supported and work in the long wavelength approximation.
Plugging the Taylor expansion of the gravitational field
\begin{equation}
h^{\mu \nu}\left(t,\mathbf{x}\right)
=\sum_{n=0}^{\infty}\,\frac{1}{n!}\,x^N\left(\partial_N h^{\mu \nu}\right)\left(t,\mathbf{0}\right)\,,
\end{equation}
into the source term of the gravitational action, the latter gives
\begin{align}\label{newsource}
S_{\text{source}}
= & \nm
-\frac{1}{2 \mpl}\,\int\!\!\dd t \int\!\! \dd^d \mathbf{x}\, T^{\mu \nu}\left(t,\mathbf{x}\right)h_{\mu \nu}\left(t,\mathbf{x}\right)\\
= & \nm
-\frac{1}{2 \mpl}\,\int\!\!\dd t\int\!\!\dd^d\mathbf{x}\,T^{\mu \nu}\left(t,\mathbf{x}\right)\sum_{n=0}^{\infty}\,\frac{1}{n!}\,x^N\partial_N h_{\mu \nu}\\
= & \nm
-\frac{1}{2 \mpl}\int\!\! \dd t\sum_{n=0}^{\infty}\,\frac{1}{n!}\left(\int\!\! \dd^d\mathbf{x}\, T^{00}x^N \right)\partial_N h_{00}-\frac{1}{\mpl}\int\!\! \dd t\sum_{n=0}^{\infty}\,\frac{1}{n!}\left(\int\!\! \dd^d\mathbf{x}\, T^{0a}x^N \right)\partial_N h_{0a}\\
&
- \frac{1}{2\mpl}\int\!\! \dd t\sum_{n=0}^{\infty}\,\frac{1}{n!}\left(\int\!\! \dd^d\mathbf{x}\, T^{ab}x^N \right)\partial_N h_{ab}\,.
\end{align}
Just as in the electromagnetic case, the action requires further partition in the conserved sectors and radiative ones.
For the purpose of expressing the radiative sector of the source action in terms of the propagating degrees of freedom $E_{ab}$, $B_{a\vert bc}$ and $\mathcal{W}_{abcd}$, we investigate the couplings to $h_{00}$, $h_{0a}$ and $h_{ab}$ in ~\eqref{newsource} separately. 

We start with the $h_{00}$ part of the action,
\begin{align}
S^{h_{00}}_{\text{source}}
= & \nm
-\frac{1}{2 \mpl}\int\!\! \dd t\sum_{n=0}^{\infty}\,\frac{1}{n!}\left(\int\!\! \dd^d\mathbf{x}\, T^{00}x^N \right)\partial_N h_{00}\\
= & \nm
-\frac{1}{2 \mpl}\int\!\! \dd t\left(\int\!\! \dd^d\mathbf{x}\, T^{00} \right) h_{00}-\frac{1}{2 \mpl}\int\!\! \dd t\left(\int\!\! \dd^d\mathbf{x}\, T^{00} x^a \right) \partial_a h_{00}\\
& 
- \frac{1}{2 \mpl}\int\!\! \dd t\sum_{n=2}^{\infty}\,\frac{1}{n!}\left(\int\!\! \dd^d\mathbf{x}\, T^{00}x^N \right)\partial_N h_{00}\,,
\end{align}
where $\mathrm{M} \equiv \int\!\! \dd^d\mathbf{x}\, T^{00}$ is the total energy of the source, and $\mathrm{G}^a\equiv\left(\int\!\! \dd^d\mathbf{x}\, T^{00} x^a \right)/\mathrm{M}$ is the center of mass (CoM) position.

Just like in the electromagnetic framework, the coefficients coupling to $h_{0a}$ can be additionally broken down into their irreducible representation \emph{via} the use of Young tableaux symmetrizers and substituting $J^a$ with $T^{0a}$. We further consider~\eqref{Tcons4}, thus we can write
\begin{align}
S^{h_{0a}}_{\text{source}}
= & \nm
-\frac{1}{\mpl}\int\!\! \dd t \left(\int\!\! \dd^d\mathbf{x}\, T^{0a} \right) h_{0a}-\frac{1}{\mpl}\int\!\! \dd t\left(\int\!\! \dd^d\mathbf{x}\, T^{0a}x^b \right)\partial_b h_{0a}\\
& \nm
- \frac{1}{\mpl}\int\!\! \dd t\sum_{n=2}^{\infty}\,\frac{1}{n!}\left(\int\!\! \dd^d\mathbf{x}\, T^{0a}x^N \right)\partial_N h_{0a}\\
= & \nm
-\frac{1}{\mpl}\int\!\! \dd t \left(\int\!\! \dd^d\mathbf{x}\, T^{0a} \right) h_{0a}-\frac{1}{\mpl}\int\!\! \dd t\left(\int\!\! \dd^d\mathbf{x}\, T^{0[a}x^{b]} \right)\partial_b h_{0a}\\
& \nm
-\frac{1}{\mpl}\int\!\! \dd t\sum_{n=1}^{\infty}\,\frac{1}{n!}\left(\int\!\! \dd^d\mathbf{x}\, T^{0(a}x^{N)} \right)\partial_N h_{0a}\\
& \nm
-\frac{1}{\mpl}\int\!\! \dd t\sum_{n=1}^{\infty}\,\frac{2n}{\left(n+1\right)!}\left(\int\!\! \dd^d\mathbf{x}\, T^{0[a}x^{i_{n}]N-1} \right)\partial_N h_{0a}\\
= & \nm
-\frac{1}{\mpl}\int\!\! \dd t \left(\int\!\! \dd^d\mathbf{x}\, T^{0a} \right) h_{0a}-\frac{1}{2 \mpl}\int\!\! \dd t\left[\int\!\! \dd^d\mathbf{x}\, \left(T^{0a}x^b-T^{0b}x^a\right) \right]\partial_b h_{0a}\\
& \nm
+\frac{1}{2 \mpl}\int\!\! \dd t\sum_{n=2}^{\infty}\,\frac{1}{n!}\left[\int\!\! \dd^d\mathbf{x}\, T^{00}x^N \right]\partial_{N-2}\left( \partial_{i_{n-1}}\partial_t h_{0 i_n }+\partial_{i_n}\partial_t h_{0 i_{n-1}}\right)\\
& 
-\frac{1}{\mpl}\int\!\! \dd t\sum_{n=2}^{\infty}\,\frac{2 n}{\left(n+1\right)!}\left[\int\!\! \dd^d\mathbf{x}\, T^{0a}x^N \right]\partial_{N-1}\left(\partial_{i_n} h_{0 a }-\partial_a h_{0 i_{n}}\right)\,.
\end{align}
where the first two terms in the last equality contains the coupling to the conserved quantities, the total linear momentum, $\mathrm{P}^a=\int\!\! \dd^d\mathbf{x}\, T^{0a}$, and the total angular momentum, $\mathrm{L}^{ab}=\int\!\! \dd^d\mathbf{x}\, \left(T^{0a}x^b-T^{0b}x^a\right)$.

Finally, the decomposition \emph{via} Young symmetrizers (once again here denoted as Young tableaux) for coefficients coupling to $h_{ab}$ yields~\cite{james1987representation,ma2004problems,Bekaert:2006py}
\begin{align}
\int\!\!\dd^d\mathbf{x}\,T^{ab}x^N = &
\frac{1}{\left(n+2\right)!}\,
\ytableausetup{mathmode, boxframe=normal, boxsize=1.67em}
\begin{ytableau}[]
\text{\footnotesize{$a$}} & \text{\footnotesize{$b$}} & \text{\footnotesize{$i_1$}} & \text{\footnotesize{$\ldots$}} & \text{\footnotesize{$i_n$}} 
\end{ytableau}
+
\frac{n+1}{\left(n+2\right)!}
\,\left(\,\,\begin{ytableau}[]
\text{\footnotesize{$a$}} & \text{\footnotesize{$b$}} & \text{\footnotesize{$i_1$}}  & \text{\footnotesize{$\ldots$}} & \text{\footnotesize{$i_{n-1}$}} \\
\text{\footnotesize{$i_n$}} & \none & \none & \none
\end{ytableau}+\text{$i$-perms}\right)\nonumber\\
& +\frac{n-1}{\left(n+1\right)!}
\,\left(\,\,\begin{ytableau}[]
\text{\footnotesize{$a$}} & \text{\footnotesize{$b$}} & \text{\footnotesize{$i_1$}}  & \text{\footnotesize{$\ldots$}} & \text{\footnotesize{$i_{n-2}$}} \\
\text{\footnotesize{$i_n$}} & \text{\footnotesize{$i_{n-1}$}} & \none & \none
\end{ytableau}+\text{$i$-perms}\right)\nonumber\\
=&
\int\!\!\dd^d\mathbf{x}\,T^{(ab}x^{N)}+\frac{4 \left(n+1\right)}{n+2}\,\underset{ab}{\mathcal{S}}\,\underset{N}{\mathcal{S}}\,\underset{a i_{n}}{\mathcal{A}}\left(\int\!\!\dd^d\mathbf{x}\,T^{a(b}x^{N)}\right)\nonumber\\
&+\frac{4 \left(n-1\right)}{n+1}\,\underset{N}{\mathcal{S}}\,\underset{ai_n}{\mathcal{A}}\,\underset{b i_{n-1}}{\mathcal{A}}\left(\int\!\!\dd^d\mathbf{x}\,T^{ab}x^{N}\right)\,.
\end{align}

Therefore, with the additional help of~\eqref{Tcons5} and~\eqref{Tcons6}, the $h_{ab}$ term in the action reads
\begin{align}
S^{h_{ab}}_{\text{source}}
= & \nm
-\frac{1}{2\mpl}\int\!\! \dd t\sum_{n=0}^{\infty}\,\frac{1}{n!}\left(\int\!\! \dd^d\mathbf{x}\, T^{ab}x^N \right)\partial_N h_{ab}\\
= & \nm
-\frac{1}{2\mpl}\int\!\! \dd t\sum_{n=0}^{\infty}\,\frac{1}{n!}\left(\int\!\! \dd^d\mathbf{x}\, T^{(ab}x^{N)} \right)\partial_N h_{ab}\\
& \nm
- \frac{1}{\mpl}\int\!\! \dd t\sum_{n=1}^{\infty}\,\frac{2\left(n+1\right)}{n!\left(n+2\right)}\,\underset{a i_n}{\mathcal{A}}\left(\int\!\! \dd^d\mathbf{x}\, T^{a(b}x^{N)} \right)\partial_N h_{ab}\\
& \nm
-\frac{1}{\mpl}\int\!\! \dd t\sum_{n=2}^{\infty}\,\frac{2\left(n-1\right)}{\left(n+1\right)!}\,\underset{a i_n}{\mathcal{A}}\,\underset{b i_{n-1}}{\mathcal{A}}\left(\int\!\! \dd^d\mathbf{x}\, T^{ab}x^{N} \right)\partial_N h_{ab}\\
= & \nm
-\frac{1}{2\mpl}\int\!\! \dd t\sum_{n=2}^{\infty}\,\frac{1}{n!}\left(\int\!\! \dd^d\mathbf{x}\, T^{00}x^{N} \right)\partial_{N-2} \partial^2_t h_{i_n i_{n-1}}\\
& \nm
- \frac{1}{\mpl}\int\!\! \dd t\sum_{n=2}^{\infty}\,\frac{2 n}{\left(n+1\right)!}\left[\int\!\! \dd^d\mathbf{x}\, T^{0a}x^N \right]\partial_{N-2}\left( \partial_{i_n}\partial_{t} h_{i_{n-1} a }-\partial_a \partial_{t} h_{i_{n-1} i_{n}}\right)\\
& 
+ \int\!\! \dd t\sum_{n=2}^{\infty}\,\frac{ n-1}{\left(n+1\right)!}\left(\int\!\! \dd^d\mathbf{x}\, T^{ab}x^N \right)\partial_{N-2}\mathcal{W}_{ai_{n}bi_{n-1}}\\
& \nm
+ \int\!\! \dd t\sum_{n=2}^{\infty}\,\frac{ n-1}{\left(n+1\right)!\left(d-2\right)}\left[\int\!\! \dd^d\mathbf{x}\, \left(T^{aa}\,x^N +T^{i_n i_{n-1}}x^{N-2}r^2-2 T^{ai_n}x^{aN-1}\right) \right]\partial_{N-2}E_{i_{n-1}i_n}\,.
\end{align}
Note that in this derivation, the coefficients carrying antisymmetrization operators over group of indices $\{a, i_n\}$ and $\{b, i_{n-1}\}$ yielded couplings to the purely spatial Riemann tensor, which in turn is replaced by its traceless counterparts using ~\eqref{expWeil}.

Adding all the components together, we write the source action~\eqref{newsource} as
\begin{equation}
S_{\text{source}}=S^{\text{cons}}_{\text{source}}+S^{\text{rad}}_{\text{source}}\,,
\end{equation}
where
\begin{subequations}
\begin{align}
S^{\text{cons}}_{\text{source}}
= & \nm
-\frac{1}{2\mpl}\int\!\! \dd t \left(\mathrm{M} h_{00}+\mathrm{M} \, \mathrm{G}^a \partial_a h_{00}+2\mathrm{P}^ah_{0a}+\mathrm{L}^{ab}\partial_ah_{0b}\right)\,,\\
\label{radiaGR}
S^{\text{rad}}_{\text{source}}
&= 
\int\!\! \dd t\sum_{n=2}^{\infty}\,\frac{1}{n!}\left(\int\!\! \dd^d\mathbf{x}\, T^{00}x^N \right)\partial_{N-2}E_{i_{n-1}i_n}\\
&\nm
+
\int\!\! \dd t\sum_{n=2}^{\infty}\,\frac{ n-1}{\left(n+1\right)!\left(d-2\right)}\left[\int\!\! \dd^d\mathbf{x}\, \left(T^{aa}\,x^N +T^{i_n i_{n-1}}x^{N-2}r^2-2 T^{ai_n}x^{aN-1}\right) \right]\partial_{N-2}E_{i_{n-1}i_n}\\
&\nm
+\int\!\! \dd t\sum_{n=2}^{\infty}\,\frac{2 n}{\left(n+1\right)!}\left(\int\!\! \dd^d\mathbf{x}\, T^{0a}x^N \right)\partial_{N-2}B_{a|i_{n-1} i_n}\\
&
+\int\!\! \dd t\sum_{n=2}^{\infty}\,\frac{ n-1}{\left(n+1\right)!}\left(\int\!\! \dd^d\mathbf{x}\, T^{ab}x^N \right)\partial_{N-2}\mathcal{W}_{ai_{n}bi_{n-1}}\,.
\end{align}
\end{subequations}
The radiative sector is expressed only in terms of couplings to propagating degrees of freedom, and the multipolar structure manifests.

Before turning to the reduction of those multipole moments as irreducible representations of SO($d$), let us confirm the results so far at the three-dimensional limit.
The conservative part of source\footnote{The orbital angular momentum vector $\mathrm{L}^a$ is recovered \emph{via} $\mathrm{L}^{ab}=\varepsilon^{abc}\mathrm{L}^c$.} and electric sectors are trivially in perfect agreement with the known three-dimensional multipolar expansion, see \emph{e.g.} Eqs.~(78) and~(79) of~\cite{Ross:2012fc}.
As for the magnetic sector, by employing the three-dimensional limit of the magnetic field ~\eqref{grEBlimit}, it becomes
\begin{align}
\lim_{d\to 3} S^{B_{a\vert bc}}_\text{source} 
= & \nm
\lim_{d\to 3} \,\int\!\! \dd t\sum_{n=2}^{\infty}\,\frac{2 n}{\left(n+1\right)!}
\left(\int\!\! \dd^d\mathbf{x}\, T^{0a}x^N \right)\partial_{N-2}B_{a|i_{n-1} i_n} \\
= & 
-\int\!\! \dd t\sum_{n=2}^{\infty}\,\frac{2 n}{\left(n+1\right)!}\left(\int\!\! \dd^d\mathbf{x}\, \varepsilon^{i_n b a} T^{0a}x^{bN-1} \right)\partial_{N-2}\,B_{ i_{n-1} i_n}\,,
\end{align}
in full agreement with the three-dimensional result, Eq.~(79) of~\cite{Ross:2012fc}.
Finally, due to the vanishing of the Weyl tensor in three dimensions, the last term of the radiative action~\eqref{radiaGR} is not relevant in such limit.

\subsection{Irreducible decomposition of the moments}\label{sec:Grav_decomp}

The last step is to rewrite the moments in ~\eqref{radiaGR} in terms of the irreducible representations of SO($d$).
Similar to the electromagnetic case, we treat the different components of $T^{\alpha \beta}$ separately depending on their tensorial nature.
We present in the main text the procedure followed to reduce the purely symmetric structure $x^L$ in ~\eqref{radiaGR} to the STF counterpart $\hat{x}^L$, and refer the interested reader to App.~\ref{sec:app_LG} for the technical details of the remaining computation regarding the complete reduction of the moments. However, we hearby remind them that identities~\eqref{grTraces}, ~\eqref{grMaxlatin1} and ~\eqref{grMaxlatin2}, along with the equations of motion, were extensively used. Additionally, we introduce hereafter the following notations for some reoccurring factor combinations 
\begin{subequations}\label{eq:shortcutsabg}
\begin{align}
&
\alj \equiv \left(\ell+2j+1\right)\left(\ell+2j+2\right)\,\ell !\,,\\
&
\blj\equiv \left(\ell+2j+1\right)\left(\ell+2j+3\right)\,\ell !\,,\\
&
\glj \equiv \left(d-2\right) \left(\ell+2j+2\right)\left(\ell+2j+3\right)\,\ell !\,,
\end{align}
\end{subequations}
together with the contractions
\begin{equation}\label{eq:shortcutsTx}
\tilde{T}^0 \equiv T^{0a}x^a\,, \qquad
\tilde{T}^a \equiv T^{ab}x^b \qquad \text{and}\qquad
\tilde{T} \equiv T^{ab}x^{ab}\,.
\end{equation}

\subsubsection{Scalar sector}\label{sec:Grav_decomp_scal}

We start with the scalar sector of the radiative action, namely the parts of the action~\eqref{radiaGR} involving $T^{00}$ and $T^{aa}$.
These terms are already symmetric in the indices, thus we only need to implement the STF relations~\eqref{xSTF2} and~\eqref{eq:xLdL}.
The $T^{00}$ piece then becomes 
\begin{align}
S^{T^{00}}_{\text{rad}}
= & \nm
\int\!\! \dd t\sum_{n=2}^{\infty}\,\frac{1}{n!}\left(\int\!\! \dd^d\mathbf{x}\, T^{00}x^N \right)\partial_{N-2}E_{i_{n-1}i_n}\\
= & \nm 
\int\!\!\dd t\,\sum_{\ell,\,j=0}^{\infty}\,\frac{\Ldlj}{\alj}\,\int\!\!\dd^d\mathbf{x}\,\partial_t^{2j}T^{00}x^{ab}\hat{x}^L r^{2j}\,\hat{\partial}_L E_{ab}\\
= & \nm
\int\!\!\dd t\,\sum_{\ell,\,j=0}^{\infty}\,\frac{\Ldlj}{\alj}\,\int\!\!\dd^d\mathbf{x}\,\partial_t^{2j}T^{00}\hat{x}^{abL} r^{2j}\,\hat{\partial}_L E_{ab}\\
& \nm
+ \int\!\!\dd t\,\sum_{\ell,\,j=0}^{\infty}\,\frac{\Ldlj\,\left(\ell+1\right)}{\alj\,\left(d+2\ell\right)}\,\int\!\!\dd^d\mathbf{x}\,\partial_t^{2j}T^{00}\,\delta^{b\langle a}\,\hat{x}^{L\rangle} r^{2j+2}\,\hat{\partial}_L E_{ab}\\
& \nm
+\int\!\!\dd t\,\sum_{\ell=1}^{\infty}\,\sum_{j=0}^{\infty}\,\frac{\Ldlj\,\ell}{\alj\,\left( d+2\ell-2\right)}\,\int\!\!\dd^d\mathbf{x}\,\partial_t^{2j}T^{00}\hat{x}^{bL-1} r^{2j+2}\,\hat{\partial}_{aL-1} E_{ab}\\
& 
+ \int\!\!\dd t\,\sum_{\ell=2}^{\infty}\,\sum_{j=0}^{\infty}\,\frac{\Ldlj\,\ell\left(\ell-1\right)}{\alj\,\left( d+2\ell-2\right)\left( d+2\ell-4\right)}\,\int\!\!\dd^d\mathbf{x}\,\partial_t^{2j}T^{00}\hat{x}^{L-2} r^{2j+4}\,\hat{\partial}_{abL-2} E_{ab}\,.
\end{align}
Using the identity in ~\eqref{delta1}, the $T^{00}$ piece can be further 
\begin{equation}\label{ST00}
S^{T^{00}}_{\text{rad}}
=
\int\!\!\dd t\,\sum_{\ell=2}^{\infty}\,\sum_{j=0}^{\infty}\,\frac{\Ldlj}{\almj{2}}\,\int\!\!\dd^d\mathbf{x}\,\partial_t^{2j}T^{00}\hat{x}^{abL-2} r^{2j}\,\hat{\partial}_{L-2} E_{ab}\,,
\end{equation}
which is explicitly in the irreducible STF form.
Similarly the $T^{aa}$ term in the action which given by
\begin{equation}
S^{T^{aa}}_{\text{rad}}
=
\frac{1}{d-2}\,\int\!\! \dd t\sum_{n=2}^{\infty}\,\frac{n-1}{(n+1)!}\left(\int\!\! \dd^d\mathbf{x}\, T^{aa}\,x^N \right)\partial_{N-2}E_{i_{n-1}i_n}\,,
\end{equation}
can be rewritten in the STF form as
\begin{equation}\label{STpp}
S^{T^{aa}}_{\text{rad}}
= 
\int\!\!\dd t\,\sum_{\ell=2}^{\infty}\,\sum_{j=0}^{\infty}\,\frac{\Ldlj}{\glmj{2}}\,\int\!\!\dd^d\mathbf{x}\,\partial_t^{2j}T^{aa}\,\hat{x}^{abL-2} r^{2j}\,\hat{\partial}_{L-2} E_{ab}\,,
\end{equation}
following the same procedure.

\subsubsection{Vector sector}\label{sec:Grav_decomp_vec}

We now move on to the vector sector, namely the $T^{0a}$ and $\tilde{T}^a = T^{ab}x^b$ terms.
First the $T^{0a}$ terms can be written as
\begin{align}\label{ST0a1}
S^{T^{0a}}_{\text{rad}}
= & \nm
\int\!\! \dd t\sum_{n=2}^{\infty}\,\frac{2 n}{\left(n+1\right)!}\left(\int\!\! \dd^d\mathbf{x}\, T^{0a}x^N \right)\partial_{N-2}B_{a|i_{n-1} i_n}\\
= & \nm \,
2\int\!\!\dd t\,\sum_{\ell,\,j=0}^{\infty}\,\frac{\Ldlj}{\blj}\,\int\!\!\dd^d\mathbf{x}\,\partial_t^{2j}T^{0a}x^{bc}\hat{x}^L r^{2j}\,\hat{\partial}_L B_{a|bc}\\
= & \nm \,
2\int\!\!\dd t\,\sum_{\ell,\,j=0}^{\infty}\,\frac{\Ldlj}{\blj}\,\int\!\!\dd^d\mathbf{x}\,\partial_t^{2j}T^{0a}\hat{x}^{bcL} r^{2j}\,\hat{\partial}_L B_{a|bc}\\
& \nm
+ 2\int\!\!\dd t\,\sum_{\ell,\,j=0}^{\infty}\,\frac{\Ldlj\,\left(\ell+1\right)}{\blj\,\left( d+2\ell\right)}\,\int\!\!\dd^d\mathbf{x}\,\partial_t^{2j}T^{0a}\,\delta^{b\langle c}\,\hat{x}^{L\rangle} r^{2j+2}\,\hat{\partial}_L B_{a|bc}\\
& \nm
+ 2\int\!\!\dd t\,\sum_{\ell=1}^{\infty}\,\sum_{j=0}^{\infty}\,\frac{\Ldlj\,\ell}{\blj\,\left( d+2\ell-2\right)}\,\int\!\!\dd^d\mathbf{x}\,\partial_t^{2j}T^{0a}\hat{x}^{bL-1} r^{2j+2}\,\hat{\partial}_{cL-1} B_{a|bc}\\
& \nm
+ 2\int\!\!\dd t\,\sum_{\ell=2}^{\infty}\,\sum_{j=0}^{\infty}\,\frac{\Ldlj\,\ell\left(\ell-1\right)}{\blj\,\left( d+2\ell-2\right)\left( d+2\ell-4\right)}\,\int\!\!\dd^d\mathbf{x}\,\partial_t^{2j}T^{0a}\hat{x}^{L-2} r^{2j+4}\,\hat{\partial}_{bcL-2} B_{a|bc}\\
= & \nm \,
2\int\!\!\dd t\,\sum_{\ell=2}^{\infty}\,\sum_{j=0}^{\infty}\,\frac{\Ldlj}{\blmj{2}}\int\!\!\dd^d\mathbf{x}\,\partial_t^{2j}T^{0a}\hat{x}^{bcL-2} r^{2j}\,\hat{\partial}_{L-2} B_{a|bc}\\
& 
+ 2\int\!\!\dd t\,\sum_{\ell=2}^{\infty}\,\sum_{j=0}^{\infty}\,\frac{\Ldlj\,\left(\ell-1\right)}{\blmj{1}\,\left(d+2\ell-2\right)}\,\int\!\!\dd^d\mathbf{x}\,\partial_t^{2j+1}T^{0a}\hat{x}^{bL-2} r^{2j}\,\hat{\partial}_{L-2} E_{ab}\,,
\end{align}
where in the last equality we apply~\eqref{delta2} and~\eqref{delta3} to rearrange the indices by symmetry. Notice the similarity with the vector sector ~\eqref{JbFIRST} in the electromagnetic case.
After some manipulations we arrive to the irreducible decomposition of the $T^{0a}$ part of the radiative action
\begin{align}\label{ST0a}
S^{T^{0a}}_{\text{rad}}
= &  \nm\,
2\int\!\!\dd t\,\sum_{\ell=2}^{\infty}\,\sum_{j=0}^{\infty}\,\frac{\Ldlj\,\left(\ell-1\right)}{\blmj{1}\,\left(d+\ell-2\right)}\,\int\!\!\dd^d\mathbf{x}\,\partial_t^{2j+1}T^{0\langle a}\hat{x}^{bL-2\rangle} r^{2j+2}\,\hat{\partial}_{L-2} E_{ab}\\
& \nm 
- 2\int\!\!\dd t\,\sum_{\ell=2}^{\infty}\,\sum_{j=0}^{\infty}\,\frac{\Ldlj\,\left(\ell-1\right)}{\blmj{1}\,\left(d+\ell-2\right)}\,\int\!\!\dd^d\mathbf{x}\,\partial_t^{2j+1}\tilde{T}^{0}\hat{x}^{abL-2} r^{2j}\,\hat{\partial}_{L-2} E_{ab}\\
& 
+ 2\int\!\!\dd t\,\sum_{\ell=2}^{\infty}\,\sum_{j=0}^{\infty}\,\frac{\Ldlj\,\ell\left(\ell-1\right)}{\ell!\,\left(\ell+2j-1\right)\left(\ell+1\right)}\left[\int\!\!\dd^d\mathbf{x}\,\partial_t^{2j}T^{0 a}\hat{x}^{bcL-2} r^{2j}\right]^{\text{TF}}\hat{\partial}_{L-2} B_{a|bc}\,.
\end{align}
%
We proceed in the same way for the $\tilde{T}^a$ piece of the action, which can be written as
\begin{align}\label{STa1}
S^{\tilde{T}^{a}}_{\text{rad}}
= &  \nm
-\frac{2}{d-2}\int\!\! \dd t\sum_{n=2}^{\infty}\,\frac{n-1}{\left(n+1\right)!}\left(\int\!\! \dd^d\mathbf{x}\, \tilde{T}^{i_n}x^{N-1} \right)\partial_{N-2}E_{i_{n-1}i_n}\\
= & \nm
-2\int\!\!\dd t\,\sum_{\ell,\,j=0}^{\infty}\,\frac{\Ldlj}{\glj}\,\int\!\!\dd^d\mathbf{x}\,\partial_t^{2j}\tilde{T}^{a}x^{b}\hat{x}^L r^{2j}\,\hat{\partial}_L E_{ab}\\
= & \nm
-2\int\!\!\dd t\,\sum_{\ell,\,j=0}^{\infty}\,\frac{\Ldlj}{\glj}\,\int\!\!\dd^d\mathbf{x}\,\partial_t^{2j}\tilde{T}^{a}\hat{x}^{bL} r^{2j}\,\hat{\partial}_L E_{ab}\\
& \nm
- 2\int\!\!\dd t\,\sum_{\ell=1}^{\infty}\,\sum_{j=0}^{\infty}\,\,\frac{\Ldlj\,\ell}{\glj\,\left( d+2\ell-2\right)}\,\int\!\!\dd^d\mathbf{x}\,\partial_t^{2j}\tilde{T}^{a}\hat{x}^{L-1} r^{2j+2}\,\hat{\partial}_{bL} E_{ab}\\
= & 
-2\int\!\!\dd t\,\sum_{\ell=1}^{\infty}\,\sum_{j=0}^{\infty}\,\frac{\Ldlj}{\glmj{1}}\,\int\!\!\dd^d\mathbf{x}\,\partial_t^{2j}\tilde{T}^{a}\hat{x}^{bL-1} r^{2j}\,\hat{\partial}_{L-1} E_{ab}\,,
\end{align}
and the final result is given by
\begin{align}\label{STa}
S^{\tilde{T}^{a}}_{\text{rad}}
= & \nm
-2\int\!\!\dd t\,\sum_{\ell=2}^{\infty}\,\sum_{j=0}^{\infty}\,\frac{\Ldlj}{\glmj{2}}\left(1+\frac{2j}{d+\ell-2}\right)\,\int\!\!\dd^d\mathbf{x}\,\partial_t^{2j}\tilde{T}^{\langle a}\hat{x}^{bL-2\rangle} r^{2j}\,\hat{\partial}_{L-2} E_{ab}\\
& \nm
+ 4\int\!\!\dd t\,\sum_{\ell=2}^{\infty}\,\sum_{j=0}^{\infty}\,\frac{\Ldlj\,j}{\glmj{2}\,\left(d+\ell-2\right)}\,\int\!\!\dd^d\mathbf{x}\,\partial_t^{2j}\tilde{T}\hat{x}^{abL-2} r^{2j-2}\,\hat{\partial}_{L-2} E_{ab}\\
& 
- 2\int\!\!\dd t\,\sum_{\ell=2}^{\infty}\,\sum_{j=0}^{\infty}\,\frac{\Ldlj\,\left(\ell-1\right)}{\glmj{1}\,\left(\ell+1\right)}\left[\int\!\!\dd^d\mathbf{x}\,\partial_t^{2j+1}\tilde{T}^{a}\hat{x}^{bcL-2} r^{2j}\right]^{\text{TF}}\hat{\partial}_{L-2} B_{a|bc}\,,
\end{align}
where the technical details of the computation for the $T^{0a}$ and $\tilde{T}^a$ terms can be found in App.~\ref{sec:app_vector}. 

\subsubsection{Tensor sector}\label{sec:Grav_decomp_tens}

Finally, the remaining $T^{ab}$ terms in the action~\eqref{radiaGR} are given by
\begin{align}
S^{T^{ab}}_{\text{rad}}
= &\nm
\frac{1}{d-2}\int\!\! \dd t\sum_{n=2}^{\infty}\,\frac{n-1}{\left(n+1\right)!}\left(\int\!\! \dd^d\mathbf{x}\,T^{i_n i_{n-1}}x^{N-2}r^2 \right)\partial_{N-2}E_{i_{n-1} i_n}\\
& 
+ \int\!\! \dd t\sum_{n=2}^{\infty}\,\frac{ n-1}{\left(n+1\right)!}\left(\int\!\! \dd^d\mathbf{x}\, T^{ab}x^N \right)\partial_{N-2}\mathcal{W}_{ai_{n}bi_{n-1}}\,.
\end{align}
This tensor sector is unique to the case of linearized gravity and has no electromagnetic equivalent.
Plugging in the STF relations~\eqref{xSTF2} and~\eqref{eq:xLdL}, the $T^{ab}$ terms can be rewritten as
\begin{align}
\label{STab}
S^{T^{ab}}_{\text{rad}}
= & \nm
\int\!\!\dd t\,\sum_{\ell,\,j=0}^{\infty}\,\frac{\Ldlj}{\glj}\,\int\!\!\dd^d\mathbf{x}\,\partial_t^{2j}T^{ab}\hat{x}^L r^{2j+2}\,\hat{\partial}_L E_{ab}\\
& \nm
+ \int\!\!\dd t\,\sum_{\ell,\,j=0}^{\infty}\,\frac{\Ldlj\,\left(d-2\right)}{\glj}\,\int\!\!\dd^d\mathbf{x}\,\partial_t^{2j}T^{ab}x^{cd}\hat{x}^L r^{2j}\,\hat{\partial}_L \mathcal{W}_{acbd}\\
= & \nm
\int\!\!\dd t\,\sum_{\ell,\,j=0}^{\infty}\,\frac{\Ldlj}{\glj}\,\int\!\!\dd^d\mathbf{x}\,\partial_t^{2j}T^{ab}\hat{x}^L r^{2j+2}\,\hat{\partial}_L E_{ab}\\
& \nm
+ \int\!\!\dd t\,\sum_{\ell,\,j=0}^{\infty}\,\frac{\Ldlj\,\left(d-2\right)}{\glj}\,\int\!\!\dd^d\mathbf{x}\,\partial_t^{2j}T^{ab}\hat{x}^{cdL} r^{2j}\,\hat{\partial}_L \mathcal{W}_{acbd}\\
& \nm
+ \int\!\!\dd t\,\sum_{\ell,\,j=0}^{\infty}\,\frac{\Ldlj\,\left(\ell+1\right)\left(d-2\right)}{\glj\,\left(d+2\ell\right)}\,\int\!\!\dd^d\mathbf{x}\,\partial_t^{2j}T^{ab}\,\delta^{d\langle c}\,\hat{x}^{L\rangle} r^{2j+2}\,\hat{\partial}_L \mathcal{W}_{acbd}\\
& \nm
+ \int\!\!\dd t\,\sum_{\ell=1}^{\infty}\,\sum_{j=0}^{\infty}\,\frac{\Ldlj\,\ell\left(d-2\right)}{\glj\,\left(d+2\ell-2\right)}\,\int\!\!\dd^d\mathbf{x}\,\partial_t^{2j}T^{ab}\hat{x}^{dL-1} r^{2j+2}\,\hat{\partial}_{cL-1} \mathcal{W}_{acbd}\\
& \nm
+ \int\!\!\dd t\,\sum_{\ell=2}^{\infty}\,\sum_{j=0}^{\infty}\,\frac{\Ldlj\,\ell\left(\ell-1\right)\left(d-2\right)}{\glj\,\left( d+2\ell-2\right)\left( d+2\ell-4\right)}\,\int\!\!\dd^d\mathbf{x}\,\partial_t^{2j}T^{ab}\hat{x}^{L-2} r^{2j+4}\,\hat{\partial}_{cdL-2} \mathcal{W}_{acbd}\\
= & \nm 
\int\!\!\dd t\,\sum_{\ell=2}^{\infty}\,\sum_{j=0}^{\infty}\,\frac{\Ldlj\,\left(d-2\right)}{\glmj{2}}\int\!\!\dd^d\mathbf{x}\,\partial_t^{2j}T^{ab}\hat{x}^{cdL-2} r^{2j}\,\hat{\partial}_{L-2} \mathcal{W}_{acbd}\\
& \nm
- 2\int\!\!\dd t\,\sum_{\ell=1}^{\infty}\,\sum_{j=0}^{\infty}\,\frac{\Ldlj\,\ell\left(d-3\right)}{\glj\,\left(d+2\ell+j\right)}\,\int\!\!\dd^d\mathbf{x}\,\partial_t^{2j+1}T^{ab}\hat{x}^{cL-1} r^{2j+2}\,\hat{\partial}_{L-1} B_{c|ab}\\
& 
+ \int\!\!\dd t\,\sum_{\ell,\,j=0}^{\infty}\,\frac{\Ldlj}{\glj}\left(1+\frac{2j\left(d-3\right)}{d+2\ell+2j}\right)\int\!\!\dd^d\mathbf{x}\,\partial_t^{2j}T^{ab}\hat{x}^L r^{2j+2}\,\hat{\partial}_L E_{ab}\,,
\end{align}
where in the last equality ~\eqref{deltaWeyl} is applied to contract the Kronecker symbols.
And the final results in terms of the irreducible representations are given by
\begin{align}\label{STabfin}
S^{T \nm^{ab}}_{\text{rad}}
&= \nm
\int\!\!\dd t\,\sum_{\ell=2}^{\infty}\,\sum_{j=0}^{\infty}\,\frac{\Ldlj}{\glmj{2}}\int\!\!\dd^d\mathbf{x}\,\partial_t^{2j}T^{\langle ab}\hat{x}^{L-2\rangle} r^{2j+2}\,\hat{\partial}_{L-2} E_{ab}\\
& \nm
+ 2 \int\!\!\dd t\,\sum_{\ell=2}^{\infty}\,\sum_{j=0}^{\infty}\,\frac{\Ldlj\,j\left(d-1\right)\left(d+2\ell+2j-1\right)}{\glmj{2}\,\left(d+\ell-1\right)\left(d+\ell-2\right)}\int\!\!\dd^d\mathbf{x}\,\partial_t^{2j}T^{\langle ab}\hat{x}^{L-2\rangle} r^{2j+2}\,\hat{\partial}_{L-2} E_{ab}\\
& \nm
+4 \int\!\!\dd t\,\sum_{\ell=2}^{\infty}\,\sum_{j=0}^{\infty}\,\frac{\Ldlj\,j}{\glmj{2}\left(d+\ell-2\right)}\left(1-\frac{\left(d-1\right)\left(d+2\ell+2j-1\right)}{\left(d+\ell-1\right)}\right)\int\!\!\dd^d\mathbf{x}\,\partial_t^{2j}T^{\langle a}\hat{x}^{b L-2\rangle} r^{2j}\,\hat{\partial}_{L-2} E_{ab}\\
& \nm
+2 \int\!\!\dd t\,\sum_{\ell=2}^{\infty}\,\sum_{j=0}^{\infty}\,\frac{\Ldlj\,j \left(d+2\ell+2j-1\right)}{\glmj{2} \,\left(d+\ell-1\right)\left(d+\ell-2\right)}\,\int\!\!\dd^d\mathbf{x}\,\partial_t^{2j}T^p_{\, p}\hat{x}^{ab L-2} r^{2j}\,\hat{\partial}_{L-2} E_{ab}\\
& \nm
+ 4\int\!\!\dd t\,\sum_{\ell=2}^{\infty}\,\sum_{j=0}^{\infty}\,\frac{\Ldlj\,j\left[2\left(j-1\right)\left(d-2\right)+\left(d+2\ell\right)\left(d-3\right)\right]}{\glmj{2}\,\left(d+\ell-1 \right)\left(d+\ell-2 \right)}\,\int\!\!\dd^d\mathbf{x}\,\partial_t^{2j}\tilde{T}\hat{x}^{ab L-2} r^{2j-2}\,\hat{\partial}_{L-2} E_{ab}\\
& \nm
+ 2\int\!\!\dd t\,\sum_{\ell=2}^{\infty}\,\sum_{j=0}^{\infty}\,\frac{\Ldlj\,\ell\left( \ell+2j+2\right)\left( d-2\right)}{\glmj{1} \,\left(\ell+1\right)\left(d+\ell-1\right)}\left[\int\!\!\dd^d\mathbf{x}\,\partial_t^{2j+1}T^{a \langle b}\hat{x}^{cL-2\rangle} r^{2j+2}\right]^{\text{TF}}\hat{\partial}_{L-2} B_{a|bc}\\
& \nm
- 2\int\!\!\dd t\,\sum_{\ell=2}^{\infty}\,\sum_{j=0}^{\infty}\,\frac{\Ldlj\,\left(\ell-1\right)\left[2j\left(d-2\right)+\left(\ell+1\right)\left(d-3\right)\right]}{\glmj{1}\,\left(\ell+1\right)\left(d+\ell-1\right)}\left[\int\!\!\dd^d\mathbf{x}\,\partial_t^{2j+1}\tilde{T}^{a}\hat{x}^{b c L-2}r^{2j}\right]^{\text{TF}}\hat{\partial}_{L-2} B_{a|bc}\\
& 
+ \int\!\!\dd t\,\sum_{\ell=2}^{\infty}\,\sum_{j=0}^{\infty}\,\frac{\Ldlj\,\left(\ell+2j\right)\left(\ell+2j+1\right)\left( d-2\right)}{\glmj{2}\,\left(\ell+1 \right)\left(\ell +2\right)}\left[\int\!\!\dd^d\mathbf{x}\,\partial_t^{2j}T^{ab}\hat{x}^{cdL-2} r^{2j}\right]^{\text{TF}}\hat{\partial}_{L-2} \mathcal{W}_{acbd}\,,
\end{align}
where the details of the reduction to irreducible representations of SO($d$) is presented in App.~\ref{sec:app_tensor}.

\subsubsection{Final expressions for the moments}\label{sec:Grav_decomp_sum}

At this stage, all contributions to the radiative part of the source action are written in terms of irreducible representations of SO($d$) and we are ready to add them together. 
The lengthy expression of the final sum is presented in~\eqref{finalLG} of App.~\ref{sec:app_sum}.
Making use of the conservation laws for the stress-energy pseudo-tensor~\eqref{Tcons} to replace the coefficients involving $T^{i_{\ell}i_{\ell-1}}$, $\tilde{T}^{i_{\ell}}$, $T^{0 i_{\ell}}$ and $T^{a i_{\ell}}$, the final action can be compacted into an elegant form
\begin{align}\label{sourcemultexpGrav}
S_{\text{source}}
= & \,\nm
S_{\text{cons}}+S^{T^{00}}_{\text{rad}}+S^{T^{aa}}_{\text{rad}}+S^{T^{0a}}_{\text{rad}}+S^{\tilde{T}^a}_{\text{rad}}+S^{T^{ab}}_{\text{rad}}\\
= & \nm
-\frac{1}{2\mpl}\int\!\! \dd t \left(\mathrm{M} h_{00}+\mathrm{M}\, \mathrm{G}^a \partial_a h_{00}+2\mathrm{P}^ah_{0a}+\mathrm{L}^{ab}\partial_ah_{0b}\right)\\
& \nm
+\int\!\!\dd t \,\sum_{\ell=2}^{\infty}\,\frac{1}{\ell !}\,I^L\,\partial_{L-2}E_{i_{\ell}i_{\ell-1}}-\int\!\!\dd t \,\sum_{\ell=2}^{\infty}\,\frac{2\ell}{\left(\ell+1\right)!}\,J^{a|L}\,\partial_{L-2}B_{a\vert i_\ell i_{\ell-1}}\\
& 
+\int\!\!\dd t \,\sum_{\ell=2}^{\infty}\,\frac{\ell-1}{\left(\ell+1\right)!}\,K^{ab|L}\,\partial_{L-2}\mathcal{W}_{a i_{\ell} b i_{\ell-1}}\,,
\end{align}
with the exact expressions for the $d$-dimensional electric, magnetic and Weyl multipole moments respectively
\begin{subequations}\label{eq:momentsGrav}
\begin{align}
I^L
= & \nm 
\sum_{j=0}^{\infty}\,\frac{\Gamma\left(\frac{d}{2}+\ell\right)}{2^{2j}j!\,\Gamma\left(\frac{d}{2}+\ell+j\right)}\left(1+\frac{4j\left(d-1\right)\left(d+\ell+j-2\right)}{\left(d-2\right)\left(d+\ell-1\right)\left(d+\ell-2\right)}\right)\int\!\!\dd^d\mathbf{x}\,\partial_t^{2j}T^{00}\hat{x}^{L} r^{2j}\\
& \nm 
-\sum_{j=0}^{\infty}\,\frac{\Gamma\left(\frac{d}{2}+\ell\right)}{2^{2j}j!\,\Gamma\left(\frac{d}{2}+\ell+j\right)}\frac{2\left(d-1\right)\left(d+\ell+2j-1\right)}{\left(d-2\right)\left(d+\ell-1\right)\left(d+\ell-2\right)}\int\!\!\dd^d\mathbf{x}\,\partial_t^{2j+1}\tilde{T}^{0}\hat{x}^{L} r^{2j}\\
& \nm
+\sum_{j=0}^{\infty}\,\frac{\Gamma\left(\frac{d}{2}+\ell\right)}{2^{2j}j!\,\Gamma\left(\frac{d}{2}+\ell+j\right)}\frac{1}{(d-2)}\left(1+\frac{2j\left(d-1\right)}{\left(d+\ell-1\right)\left(d+\ell-2\right)}\right)\int\!\!\dd^d\mathbf{x}\,\partial_t^{2j}T^{aa}\,\hat{x}^{L} r^{2j}\\
&\
+\sum_{\,j=0}^{\infty}\,\frac{\Gamma\left(\frac{d}{2}+\ell\right)}{2^{2j}j!\,\Gamma\left(\frac{d}{2}+\ell+j\right)}\frac{\left(d-1\right)}{\left(d-2\right)\left(d+\ell-1\right)\left(d+\ell-2\right)} \int\!\!\dd^d\mathbf{x}\,\partial_t^{2j+2}\tilde{T}\hat{x}^{L} r^{2j}\,,\\
J^{a|L}
= &\nm
\underset{a i_{\ell}}{\mathcal{A}}\,\sum_{j=0}^{\infty}\,\frac{\Gamma\left(\frac{d}{2}+\ell\right)}{2^{2j}j!\,\Gamma\left(\frac{d}{2}+\ell+j\right)}\left(1+\frac{2j}{\left(d+\ell-1\right)}\right)\left[\int\!\!\dd^d\mathbf{x}\,\partial_t^{2j}T^{0a}\hat{x}^{L} r^{2j}\right]^{\text{TF}}\\
&
-\underset{a i_{\ell}}{\mathcal{A}}\,\sum_{j=0}^{\infty}\,\frac{\Gamma\left(\frac{d}{2}+\ell\right)}{2^{2j}j!\,\Gamma\left(\frac{d}{2}+\ell+j\right)}\frac{1}{\left( d+\ell-1\right)}\,\left[\int\!\!\dd^d\mathbf{x}\,\partial_t^{2j+1}\tilde{T}^{a}\hat{x}^{L} r^{2j}\right]^{\text{TF}}
\,,\\
K^{ab|L}
= &
\underset{a i_{\ell}}{\mathcal{A}}\,\underset{b i_{\ell-1}}{\mathcal{A}}\,\sum_{j=0}^{\infty}\,\frac{\Gamma\left(\frac{d}{2}+\ell\right)}{2^{2j}j!\,\Gamma\left(\frac{d}{2}+\ell+j\right)}\left[\int\!\!\dd^d\mathbf{x}\,\partial_t^{2j}T^{a b}\hat{x}^{L} r^{2j}\right]^{\text{TF}}\,.
\end{align}
\end{subequations}
The electric moment correspond to the symmetric Young tableau
\begin{equation}
I^L =
\ytableausetup{mathmode, boxframe=normal, boxsize=1.7em}
\begin{ytableau}[]
i_\ell & i_{\ell-1}  & \ldots & i_2 & i_1 
\end{ytableau}\,,
\end{equation}
when the two other moments are respectively given by the mixed Young tableaux~\cite{james1987representation,ma2004problems,Bekaert:2006py}
\begin{equation}
J^{a\vert L} =
\ytableausetup{mathmode, boxframe=normal, boxsize=1.7em}
\begin{ytableau}[]
i_\ell & i_{\ell-1}  & \ldots & i_2 & i_1 \\
a & \none & \none & \none & \none
\end{ytableau}
\qquad\text{and}\qquad
K^{ab\vert L} =
\ytableausetup{mathmode, boxframe=normal, boxsize=1.7em}
\begin{ytableau}[]
i_\ell & i_{\ell-1}  & i_{\ell-2}  & \ldots & i_2 & i_1 \\
a & b & \none & \none & \none & \none
\end{ytableau}\,.
\end{equation}

Note that the three-dimensional limits of the multipoles $I^L$ and (the dual of) $J^{a\vert L}$ perfectly agree with the known three-dimensional results, ~(105) and~(106) of~\cite{Ross:2012fc}, whereas the additional set of moments $K^{ab\vert L}$ is absent in three-dimensions.

\section{Conclusions}\label{sec:concl}
We have extended to a generic number of spatial dimensions the results presented in~\cite{Ross:2012fc} for a scalar field, electromagnetism and linearized gravity.
Our results confirm that electric-type moments can be readily generalized to $d$ spatial dimensions, while magnetic-type moments have to be represented by expressions having the symmetries of a mixed Young tableaux. Furthermore, within the framework of linearized gravity, we have identified a novel set of `Weyl-type' moments, with symmetries of another type of mixed Young tableaux.
These additional moments couple to the spatial Weyl tensor and are absent in three dimensions, in agreement with the discussion presented in~\cite{Henry:2021cek}, where a different formalism and gauge are considered. The expressions of the gravitational moments~\eqref{eq:momentsGrav} are crucial ingredients towards high accuracy gravitational waveforms within the EFT framework. Indeed, they are the key ingredients of the GW flux, the computation of which entails (logarithmic) divergences starting at the 3PN order. This provided our main motivation for this work, since one then needs to obtain the expression of the (source) mass quadrupole moment, $I^{ij}$, in arbitrary dimensions. The derivation of the 3PN GW flux will be discussed elsewhere. Needless to say, the results given in this work will be building blocks towards constructing accurate waveforms at even higher PN orders. To conclude, let us remark that we have excluded throughout this work the inclusion of non-linear terms in the action. Within the EFT context, the so-called ``tail-of-tail'' effects due to the gravitational interactions with the background geometry in the far zone must be taken into account in the computation of the gravitational flux starting at 3PN~\cite{Goldberger:2009qd}. Moreover, non-linear terms, incorporating notably dissipative effects, will be of prime importance when reaching 4PN, where the interplay between conservative and dissipative dynamics affects the gravitational flux~\cite{Blanchet:2023sbv}.
We reserve this exciting new avenue for future work.

\acknowledgments

It is a pleasure to thank L. Blanchet, G. Faye, Q. Henry and R. A. Porto for enlightening discussions and comments.
The work of F.L. and Z.Y. was funded by the ERC Consolidator Grant ``Precision Gravity: From the LHC to LISA'' provided by the European Research Council (ERC) under the European Union’s H2020 research and innovation program (grant agreement No. 817791). The work of L.A. was supported by the International Helmholtz-Weizmann Research School for Multimessenger Astronomy, largely funded through the Initiative and Networking Fund of the Helmholtz Association.

\appendix


\section{Formulas for irreducible tensor decomposition in $d$ dimensions}\label{sec:app_irred_decomp}

This appendix lists expressions and relations that are useful when computing the irreducible decomposition of tensors of SO($d$).\\

Arbitrary symmetric tensors $S^N$ are expressed in a STF guise as
\begin{align}
	S^{N} = \sum^{\left[n/2\right]}_{p=0} \,\frac{n!}{\left(n-2p\right)!}\,\Lambda^{(d)}_{n-2p,p}\,\delta^{(i_{1}i_{2}}...\,\delta^{i_{2p-1}i_{2p}}\,\hat{S}^{i_{2p+1}...i_{n})a_{1}a_{1}...a_{p}a_{p}}\,,
	\label{eq:STFformulaapp}
\end{align}
where $[n/2]$ denotes the integer part of $n/2$ and where we defined
\begin{equation}
\Lambda^{(d)}_{n,p} \equiv \frac{\Gamma \left( \frac{d}{2}+n\right)}{2^{2p}p!\,\Gamma \left( \frac{d}{2}+n+p\right)}\,.\label{lambdaapp}
\end{equation}
Therefore, products such as $x^N$ can be rewritten as as~\cite{Blanchet:2003gy, Blanchet:2005tk} 
\begin{equation}\label{xSTF1}
x^N=\sum_{p=0}^{\left[n/2\right]}\,\frac{n!}{\left(n-2p\right)!}\,\Lambda^{(d)}_{n-2p,p}\,\delta^{(2P}\,\hat{x}^{N-2P)}r^{2p}\,,
\end{equation}
where $\delta^{2P}$ is a product of $p$ Kronecker symbols.
With a little manipulation, this leads to the extremely useful relation
\begin{equation}\label{xSTF2}
\sum_{\ell=0}^{\infty}\,\frac{1}{\ell!}x^L\partial_L=\sum_{\ell,\, j=0}^{\infty}\,\frac{\Ldlj}{\ell!}\,r^{2j}\hat{x}^L\,\nabla^{2j}\hat{\partial}_{L}\,.
\end{equation}

Given a tensor $\mathcal{J}^{aL}$, STF in the indices $\{L\}$, and a tensor $\mathcal{T}^{abL}$, separately STF in the pair $\{a,b\}$ and the indices $\{L\}$, one can extract the symmetric and antisymmetric parts as~\cite{Henry:2021cek}
\begin{subequations}
\begin{align}
&
\mathcal{J}^{aL}=\mathcal{J}^{\left(aL \right)}+\frac{2 \ell}{\ell+1}\,\underset{ L}{\mathcal{S}}\,\mathcal{J}^{\left[ai_{\ell}\right]L-1}\,,\label{eqSymmAnti1}\\
&
\mathcal{T}^{abL}=\mathcal{T}^{\left(abL\right)}+\frac{4\left(\ell+1 \right)}{\ell+2}\,\underset{ab}{\mathcal{S}}\,\underset{L}{\mathcal{S}}\,\underset{a i_{\ell}}{\mathcal{A}}\, \mathcal{T}^{a \left(b L \right)}+\frac{4\left(\ell-1 \right)}{\ell+1}\,\underset{L}{\mathcal{S}}\,\underset{a i_{\ell}}{\mathcal{A}}\,\underset{b i_{\ell-1}}{\mathcal{A}}\, \mathcal{T}^{abL}\,.\label{eqSymmAnti2}
\end{align}
\end{subequations}
The irreducible decompositions of the same objects into their corresponding TF counterparts read~\cite{Henry:2021cek}
\begin{equation}\label{eqTFone}
\mathcal{J}^{aL}
=
\left[\mathcal{J}^{aL}\right]^{\text{TF}}
+\frac{\ell\left( d+2\ell-4\right)}{\left(d+\ell-3 \right)\left(d+2\ell-2 \right)}\,\delta^{a(i_{\ell}}\,\mathcal{Q}^{L-1)}
-\frac{\ell\left(\ell-1\right)}{\left( d+\ell-3\right)\left(d+2\ell-2 \right)}\,\delta^{(i_{\ell}i_{\ell-1}}\,\mathcal{Q}^{L-2)a}\,,
\end{equation}
and
\begin{align}\label{eqTFtwo}
\mathcal{T}^{abL}
= & \nm
\left[\mathcal{T}^{abL}\right]^{\text{TF}}\\
& \nm
+\frac{2\ell\left(d+2\ell-4\right)}{\left( d+\ell-2\right)\left(d+2\ell-2\right)}\,\underset{ab,\,L}{\mathcal{S}}\,\delta^{ai_{\ell}}\,\left[\mathcal{H}^{b L-1}+\frac{4\ell}{\left(d-2\right)\left(d+2\ell\right)}\,\mathcal{H}^{(b L-1)}\right]\\
& \nm
-\frac{2\ell \left(\ell-1\right)}{\left( d+\ell-2\right)\left(d+2\ell-2\right)}\,\underset{ab,\,L}{\mathcal{S}}\,\delta^{i_{\ell}i_{\ell-1}}\,\left[\mathcal{H}^{a bL-2}+\frac{4\ell}{\left(d-2\right)\left(d+2\ell\right)}\,\mathcal{H}^{(a bL-2)}\right]\\
& \nm
-\frac{2 \ell\left(d+2\ell-4 \right)}{\left(d-2\right)\left(d+\ell-2\right)\left(d+2\ell\right)}\,\delta^{ab}\,\mathcal{H}^{(i_{\ell} L-1)}\\
& \nm
+\frac{\ell\left(\ell-1\right)\left(d+2\ell-6\right)}{\left(d+\ell-3\right)\left(d+\ell-4\right)\left(d+2\ell-2\right)}\,\underset{L}{\mathcal{S}}\,\delta^{a i_{\ell}}\,\delta^{b i_{\ell-1}}\,\mathcal{L}^{L-2}\\
& \nm
-\frac{2\ell\left(\ell-1\right)\left(\ell-2\right)\left(d+2\ell-6\right)}{\left(d+\ell-3\right)\left(d+\ell-4\right)\left(d+2\ell-2\right)\left(d+2\ell-4\right)}\,\underset{ab}{\mathcal{S}}\,\delta^{a(i_{\ell}}\,\delta^{i_{\ell-1}i_{\ell-2}}\,\mathcal{L}^{L-3)b}\\
& \nm
+\frac{\ell\left(\ell-1\right)\left(\ell-2\right)\left(\ell-3\right)}{\left(d+\ell-3\right)\left(d+\ell-4\right)\left(d+2\ell-2\right)\left(d+2\ell-4\right)}\,\delta^{(i_{\ell}i_{\ell-1}}\,\delta^{i_{\ell-2}i_{\ell-3}}\,\mathcal{L}^{L-4)ab}\\
& 
-\frac{\ell\left(\ell-1\right)\left(d+2\ell-6\right)}{\left(d+\ell-3\right)\left(d+\ell-4\right)\left(d+2\ell-2\right)\left(d+2\ell-4\right)}\,\delta^{ab}\,\delta^{(i_{\ell} i_{\ell-1}}\mathcal{L}^{L-2)}\,,
\end{align}
where we defined the trace-free parts of the tensors as $\left[\mathcal{J}^{aL}\right]^{\text{TF}}=\underset{aL}{\text{TF}}\,\mathcal{J}^{aL}$  and $\left[\mathcal{T}^{abL}\right]^{\text{TF}}=\underset{abL}{\text{TF}}\,\mathcal{T}^{abL}$, and introduced the tensors 
\begin{equation}
\mathcal{Q}^{L-1} \equiv \mathcal{J}^{aaL-1}\,,
\qquad
\mathcal{H}^L \equiv \underset{L}{\text{TF}}\,\mathcal{T}^{ai_\ell aL-1}
\qquad\text{and}\qquad
\mathcal{L}^{L-2}\equiv\mathcal{T}^{ababL-2}\,,
\end{equation}
which are STF in all their indices.
Applying those relations to the simplest case of coordinates and derivatives, one finds the relations
\begin{subequations}\label{eq:xLdL}
\begin{align}
&
\hat{x}^L=x^{i_{\ell}}\,\hat{x}^{L-1}-\frac{\left(\ell-1\right)r^2}{d+2\ell-4}\,\delta^{i_{\ell}\langle i_{\ell-1}}\,\hat{x}^{L-2\rangle}\,,\label{OUx}\\
&
\hat{\partial}_L=\partial_{i_{\ell}}\hat{\partial}_{L-1}-\frac{\ell-1}{d+2\ell-4}\,\delta_{i_{\ell}\langle i_{\ell-1}}\hat{\partial}_{L-2\rangle}\nabla^2\,,
\label{OUdelta}
\end{align}
\end{subequations}
that are used extensively throughout this work.

\section{Conservation laws for the electromagnetic current and the stress-energy pseudo-tensor}\label{sec:app_conservation}

This appendix contains useful formulas deriving from the conservation laws of the sources.

In the case of electromagnetism described in Sec.~\ref{sec:electromag}, the conservation of the four-current $\partial_{\alpha}J^{\alpha}=0$ yields the identities (valid for any $j, \ell \geq 0$)
\begin{equation}\label{Jcons2}
\int\!\!\dd^d\mathbf{x}\,\partial_t J^0 r^{2j} x^L 
=
\int\!\!\dd^d\mathbf{x}\left(\ell J^{(i_{\ell}}x^{L-1)}r^{2j}+2j \tilde{J}x^Lr^{2j-2}\right)\,,
\end{equation}
where we recall our notation $\tilde{J} \equiv J^ax^a$.

Similarly, in the case of linearized gravity investigated in Sec.~\ref{sec:gravity}, the conservation of the stress-energy pseudo-tensor $\partial_{\alpha}T^{\alpha \beta}=0$ can be translated into a set of relations (valid for any $j, \ell \geq 0$) 
\begin{subequations}
\label{Tcons}
\begin{align}
&
\int\!\!\dd^d\mathbf{x}\,T^{(i_{\ell-1}i_{\ell}}x^{L-2)}r^{2j+2} 
=\frac{1}{\ell-1}\int\!\!\dd^d\mathbf{x}\,\partial_tT^{0 (i_{\ell}}x^{L-1)}r^{2j+2}-\frac{2\left(j+1\right)}{\ell-1}\int\!\!\dd^d\mathbf{x}\,\tilde{T}^{ (i_{\ell}}x^{L-1)}r^{2j}\,,\label{Tcons4}\\
&
\int\!\!\dd^d\mathbf{x}\,\tilde{T}^{(i_{\ell}}x^{L-1)}r^{2j}
= \frac{1}{\ell}\int\!\!\dd^d\mathbf{x}\,\partial_t\tilde{T}^{0}x^{L}r^{2j}-\frac{2j}{\ell}\int\!\!\dd^d\mathbf{x}\,\tilde{T}x^{L}r^{2j-2}-\frac{1}{\ell}\int\!\!\dd^d\mathbf{x}\,T^{aa}\,x^Lr^{2j}\,,\label{Tcons5}\\
&
\int\!\!\dd^d\mathbf{x}\,T^{0(i_{\ell}}x^{L-1)}r^{2j+2}
=\frac{1}{\ell}\int\!\!\dd^d\mathbf{x}\,\partial_tT^{00}x^{L}r^{2j+2}-\frac{2\left(j+1\right)}{\ell}\int\!\!\dd^d\mathbf{x}\,\tilde{T}^{0}x^{L}r^{2j}\,,\label{Tcons6}\\
&
\int\!\!\dd^d\mathbf{x}\,T^{a(i_{\ell}}x^{L-1)}r^{2j+2}
=\frac{1}{\ell}\int\!\!\dd^d\mathbf{x}\,\partial_t T^{0a}x^{L}r^{2j+2}-\frac{2\left(j+1\right)}{\ell}\,\int\!\!\dd^d\mathbf{x}\,\tilde{T}^{a}x^{L}r^{2j}\,,\label{Tcons7}
\end{align}
\end{subequations}
with the help of integration-by-parts. 
We remind the reader the shorthand notations introduced in ~\eqref{eq:shortcutsTx}, namely $\tilde{T}^0=  T^{0a}x^a$, $\tilde{T}^a = T^{ab}x^b$ and $\tilde{T} =  T^{ab}x^{ab}$.

Note that, although derived in $d$ dimensions, these relations are similar to the three-dimensional ones used in~\cite{Ross:2012fc}.

\section{Technical details of the irreducible decomposition}\label{sec:app_technical_details}

This appendix collects technical steps that are followed when decomposing the multipole moments into their irreducible counterparts, for both electromagnetism and linearized gravity.

\subsection{Electromagnetism}\label{sec:app_EMSTF}

Let us detail how we decomposed the electromagnetic radiative source terms ~\eqref{SJ0} and~\eqref{JbFIRST} into irreducible multipoles, as given in ~\eqref{sourcemultexpEM}.

The scalar sector, ~\eqref{SJ0}, is already in the sought form, so we will deal here with the vector one, ~\eqref{JbFIRST}, namely
\begin{align}\label{JbFIRST_app}
S_{\text{rad}}^{J^a} 
= & \nm
\int\!\!\dd t\,\sum_{\ell=1}^{\infty}\,\sum_{j=0}^{\infty}\,\frac{\Ldlj}{\left(\ell-1\right)!\,\left(\ell+2j+1\right)}\,\int\!\!\dd^d\mathbf{x}\,\partial_t^{2j}J^a\hat{x}^{bL-1}r^{2j}\,\hat{\partial}_{L-1} B_{a|b}\\
& \ 
+ \int\!\!\dd t\,\sum_{\ell=1}^{\infty}\,\sum_{j=0}^{\infty}\,\frac{\Ldlj}{\left(\ell-1\right)!\,\left(\ell+2j+2\right)\left(d+2\ell-2\right)}\,\int\!\!\dd^d\mathbf{x}\,\partial_t^{2j+1}J^a\hat{x}^{L-1}r^{2j}\,\hat{\partial}_{L-1} E^a\,.
\end{align}
The first line has nearly the appropriate symmetries for a magnetic-type moment: it is STF in its $\{b, L-1\}$ indices and antisymmetric in $\{a,b\}$, so it only requires a removal of the trace, which is easily done by applying the relation~\eqref{eqTFone}.
As for the second line, let us first symmetrise it, using the relation~\eqref{eqSymmAnti1}
\begin{align}\label{decoEM0}
\int\!\!\dd^d\mathbf{x}\,\partial_t^{2j+1}J^a\hat{x}^{L-1}r^{2j}\,\hat{\partial}_{L-1} E^a
= & \nm
\int\!\!\dd^d\mathbf{x}\,\partial_t^{2j+1}J^{(a}\hat{x}^{L-1)}r^{2j}\,\hat{\partial}_{L-1} E^a\\
& 
+\frac{2\left(\ell-1\right)}{\ell}\,\underset{L-1}{\mathcal{S}}\left(\int\!\!\dd^d\mathbf{x}\,\partial_t^{2j+1}J^{[a}\hat{x}^{i_{\ell-1}]L-2}r^{2j}\right)\hat{\partial}_{L-1} E^a\,.
\end{align}
Using the Maxwell equations~\eqref{MaxSpatial}, the relations~\eqref{eq:xLdL}, and removing the traces with the help of ~\eqref{eqTFone}, one obtains irreducible expressions for the coefficients entering both the first line of~\eqref{JbFIRST_app} and~\eqref{decoEM0} as
\begin{subequations}
\begin{align}
J^a\hat{x}^{b L-1}\,\hat{\partial}_{L-1}B_{a \vert b}
= & \nm\,
\left[J^a\hat{x}^{b L-1}\right]^{\text{TF}}\,\hat{\partial}_{L-1}B_{a|b}\\
& \nm \,
-\frac{\left(\ell-1\right)^2 r^2}{\left(d+\ell-3\right)\left(d+2\ell-4\right)}\,J^{\langle a}\hat{x}^{L-2 \rangle}\,\hat{\partial}_{L-2}\partial_t E^{a}\nonumber\\
& 
+\frac{\ell-1}{d+\ell-3}\,\tilde{J}\hat{x}^{a L-2}\,\hat{\partial}_{L-2}\partial_t E^{a}\,,\\
J^{(a}\hat{x}^{L-1)}\hat{\partial}_{L-1}E^a
= & \nm\,
J^{\langle a}\hat{x}^{L-1 \rangle}\,\hat{\partial}_{L-1}E^a \\
& \nm\,
+\frac{2 \left( \ell-1\right)\left( \ell-2\right)^2 r^2}{\ell \left(d+2\ell-4 \right)\left(d+2\ell-6 \right)^2}\,J^{\langle a}\hat{x}^{L-3 \rangle}\,\hat{\partial}_{L-3}\partial^2_t E^a\\
&
-\frac{2 \left(\ell -1\right)\left(\ell -2\right)}{\ell \left(d+2\ell-4\right)\left(d+2\ell-6\right)}\,\tilde{J}\hat{x}^{a L-3}\, \hat{\partial}_{L-3}\partial^2_t E^a\,,\\
\underset{L-1}{\mathcal{S}}\, J^{[a}\hat{x}^{i_{\ell-1}]L-2}\,\hat{\partial}_{L-1}E_{a}
= & \nm\,
\frac{1}{2}\,\left[J^a\hat{x}^{b L-2}\right]^{\text{TF}}\,\hat{\partial}_{L-2}\partial_tB_{a|b}\\
& \nm \,
-\frac{\left( \ell-2 \right)^3 r^2}{2\left(d+\ell-4\right)\left(d+2\ell-6\right)^2}\,\,J^{\langle a}\hat{x}^{L-3 \rangle}\,\hat{\partial}_{L-3}\partial^2_t E^a\\
&
+\frac{\left(\ell-2\right)^2}{2\left(d+\ell-4\right)\left(d+2\ell-6\right)}\,\tilde{J}\hat{x}^{a L-3}\,\hat{\partial}_{L-3}\partial^2_t E^a\,,
\end{align}
\end{subequations}
where $\tilde{J} \equiv J^ax^a$.
Those identities allow us to rewrite ~\eqref{JbFIRST_app} in terms of irreducible representations of SO($d$) as
\begin{align}
S_{\text{rad}}^{J^a} 
= & \nm
-\int\!\!\dd t\,\sum_{\ell=1}^{\infty}\,\sum_{j=0}^{\infty}\,\frac{\Ldlj}{\left(\ell-1\right)!\,\left(\ell+2j+2\right)\left(d+\ell-2\right)}\,\int\!\!\dd^d\mathbf{x}\,\partial_t^{2j+1}\tilde{J}\hat{x}^{aL-1}r^{2j}\,\hat{\partial}_{L-1} E^a\\
& \nm
+ \int\!\!\dd t\,\sum_{\ell=1}^{\infty}\,\sum_{j=0}^{\infty}\,\frac{\Ldlj}{\left(\ell-1\right)!\,\left(\ell+2j+2\right)\left(d+\ell-2\right)}\,\int\!\!\dd^d\mathbf{x}\,\partial_t^{2j+1}J^{\langle a}\hat{x}^{L-1\rangle}r^{2j}\,\hat{\partial}_{L-1} E^a\\
& 
  \int\!\!\dd t\,\sum_{\ell=1}^{\infty}\,\sum_{j=0}^{\infty}\,\frac{\Ldlj\,\ell}{\left(\ell+1\right)!}\left[\int\!\!\dd^d\mathbf{x}\,\partial_t^{2j}J^{a}\hat{x}^{b L-1}r^{2j}\right]^{\text{TF}}\hat{\partial}_{L-1} B_{a|b}\,.
\end{align}
Adding the scalar sector $S_{\text{rad}}^{J^0}$~\eqref{SJ0} and using the conservation of the current, ~\eqref{Jcons2}, the electromagnetic radiative action can be written as
\begin{align}
S_{\text{rad}}
= & \nm
\int\!\!\dd t\,\sum_{\ell,\,j=0}^{\infty}\,\frac{\Ldlj}{\ell!}\left(1+\frac{2j}{d+\ell-2}\right)\int\!\!\dd^d\mathbf{x}\,\partial_t^{2j}J^0\hat{x}^{aL-1}r^{2j}\,\hat{\partial}_L E^a\\
& \nm
-\int\!\!\dd t\,\sum_{\ell=1}^{\infty}\,\sum_{j=0}^{\infty}\,\frac{\Ldlj}{\ell!\,\left(d+\ell-2\right)}\,\int\!\!\dd^d\mathbf{x}\,\partial_t^{2j+1}\tilde{J}\hat{x}^{aL-1}r^{2j}\,\hat{\partial}_{L-1} E^a\\
& 
+ \int\!\!\dd t\,\sum_{\ell=1}^{\infty}\,\sum_{j=0}^{\infty}\,\frac{\Ldlj\,\ell}{\left(\ell+1\right)!}\,\left[\int\!\!\dd^d\mathbf{x}\,\partial_t^{2j}J^{a}\hat{x}^{b L-1}r^{2j}\right]^{\text{TF}}\hat{\partial}_{L-1} B_{a|b}\,.
\end{align}
This final expression directly gives the result for the irreducible decomposition of the electromagnetic action, ~\eqref{sourcemultexpEM} and~\eqref{EMMOM}.

\subsection{Linearized gravity}
\label{sec:app_LG}

Let us now turn to the case of linearized gravity, described in Sec.~\ref{sec:Grav_decomp}.
This appendix hence details the necessary steps to decompose in an irreducible fashion the radiative action ~\eqref{radiaGR}.
Hereafter, we will often use the following identities, that are consequences of the formulas exposed in App.~\ref{sec:app_irred_decomp}
\begin{subequations}
\begin{align}
&
\delta^{a\langle b}\,\hat{x}^{L\rangle}\,\hat{\partial}_L E_{ab}=
\frac{\ell \left( d+2\ell-4\right)}{\left( \ell+1\right)\left(d+2\ell-2 \right)}\,\hat{x}^{aL-1}\,\hat{\partial}_{b L-1} E_{ab}\,,\label{delta1}\\
&
\hat{x}^L\,\delta_{a\langle b}\,\hat{\partial}_{L\rangle}B_{c|ab}=\frac{\ell}{\ell+1}\,\hat{x}^{aL-1}\,\hat{\partial}_{bL-1}B_{c|ab}-\frac{2\ell}{\left( \ell+1\right)\left( d+2\ell-2\right)}\,\hat{x}^{aL-1}\,\hat{\partial}_{bL-1}B_{c|ba}\,,\label{delta2}\\
&
\hat{x}^L\,\delta_{b\langle a}\,\hat{\partial}_{L\rangle}B_{c|ab} =\frac{\ell}{\ell+1}\,\hat{x}^{aL-1}\,\hat{\partial}_{bL-1}B_{c|ba}-\frac{2\ell}{\left( \ell+1\right)\left( d+2\ell-2\right)}\,\hat{x}^{aL-1}\,\hat{\partial}_{bL-1}B_{c|ab}\,,\label{delta3}\\
&
T^{ab}\hat{x}^{cdL-2}\,\delta_{c\langle b}\,\hat{\partial}_{L-2\rangle}E_{ad}=
\frac{d-2}{\left( \ell-1\right)\left( d+2\ell-6\right)}\,T^{ab}\hat{x}^{acL-2}\,\hat{\partial}_{L-2}E_{bc}\,,\label{delta4}\\
&
T^{ab}\hat{x}^{cdL-2}\,\delta_{a\langle b}\,\hat{\partial}_{L-2\rangle}E_{cd} =
\frac{\left(\ell-2\right)\left(d+2\ell-8\right)}{\left( \ell-1\right)\left(d+2\ell-6\right)}\,T^{ab}\hat{x}^{bcdL-3}\,\hat{\partial}_{aL-3}E_{cd} \label{delta5}\\
& \nm \hspace{5cm}
+\frac{T^{aa}}{\ell-1}\,\hat{x}^{abL-2}\,\hat{\partial}_{L-2}E_{ab} \,,\\
&
T^{ab}\,\delta^{d\langle c}\hat{x}^{L \rangle}\,\hat{\partial}_L \mathcal{W}_{acbd}= \frac{\ell \left(d+2\ell-4 \right)}{\left(\ell+1 \right)\left(d+2\ell-2 \right)}\,T^{ab}\hat{x}^{dL-1}\,\hat{\partial}_{cL-1}\mathcal{W}_{acbd}\,,\label{deltaWeyl}\\
&
T^{ab}\hat{x}^{abL-2}
= \left(\tilde{T} - \frac{r^2\,T^p_{\, p}}{d+2\ell-4}\right)\hat{x}^{L-2}
- \frac{2(\ell-2)\,r^2}{d+2\ell-4}\tilde{T}^{\langle i_{\ell-2}}\hat{x}^{L-3\rangle}\\
& \nm \hspace{5cm}
+ \frac{(\ell-2)(\ell-3)\,r^4}{(d+2\ell-4)(d+2\ell-6)} T^{\langle i_{\ell-2}i_{\ell-3}}\hat{x}^{L-4\rangle}\,.
\end{align}
\end{subequations}
%

\subsubsection{Scalar sector}\label{sec:app_scalar}

The $T^{00}$ and $T^p_{\, p}$ terms are treated in the main text, in Sec.~\ref{sec:Grav_decomp_scal}. Their expressions in terms of irreducible decomposition of SO($d$) are given in \eqref{ST00} and~\eqref{STpp}, respectively.

\subsubsection{Vector sector}\label{sec:app_vector}

The $T^{0a}$ and $\tilde{T}^a$ terms are respectively given by ~\eqref{ST0a1} and~\eqref{STa1},
\begin{subequations}
\begin{align}\label{ST0a1_app}
S^{T^{0a}}_{\text{rad}}
= & \nm\,
2\int\!\!\dd t\,\sum_{\ell=2}^{\infty}\,\sum_{j=0}^{\infty}\,\frac{\Ldlj}{\blmj{2}}\int\!\!\dd^d\mathbf{x}\,\partial_t^{2j}T^{0a}\hat{x}^{bcL-2} r^{2j}\,\hat{\partial}_{L-2} B_{a|bc}\\
& 
+ 2\int\!\!\dd t\,\sum_{\ell=2}^{\infty}\,\sum_{j=0}^{\infty}\,\frac{\Ldlj\,\left(\ell-1\right)}{\blmj{1}\,\left(d+2\ell-2\right)}\,\int\!\!\dd^d\mathbf{x}\,\partial_t^{2j+1}T^{0a}\hat{x}^{bL-2} r^{2j}\,\hat{\partial}_{L-2} E_{ab}\,,\\
\label{STa1_app}
S^{\tilde{T}^{a}}_{\text{rad}}
= & 
-2\int\!\!\dd t\,\sum_{\ell=1}^{\infty}\,\sum_{j=0}^{\infty}\,\frac{\Ldlj}{\glmj{1}}\,\int\!\!\dd^d\mathbf{x}\,\partial_t^{2j}\tilde{T}^{a}\hat{x}^{bL-1} r^{2j}\,\hat{\partial}_{L-1} E_{ab}\,.
\end{align}
\end{subequations}
Exactly as is the case for electromagnetism in App.~\ref{sec:app_EMSTF}, the first line needs only its trace to be removed, whereas the two other lines require more work.
Upon using ~\eqref{eqSymmAnti1}, we can decompose
\begin{subequations}
\begin{align}
\int\!\!\dd^d\mathbf{x}\,\partial_t^{2j+1}T^{0a}\hat{x}^{bL-2} r^{2j+2}\,\hat{\partial}_{L-2} E_{ab}
= &
\int\!\!\dd^d\mathbf{x}\,\partial_t^{2j+1}T^{0(a}\hat{x}^{bL-2)} r^{2j+2}\,\hat{\partial}_{L-2} E_{ab}\\
& \nm
+\frac{2\left(\ell-1\right)}{\ell}\,\underset{bL-2}{\mathcal{S}}\left(\int\!\!\dd^d\mathbf{x}\,\partial_t^{2j+1}T^{0[a}\hat{x}^{b]L-2} r^{2j+2}\right)\hat{\partial}_{L-2} E_{ab}\,,\\
\int\!\!\dd^d\mathbf{x}\,\partial_t^{2j}\tilde{T}^{a}\hat{x}^{bL-1} r^{2j}\,\hat{\partial}_{L-1} E_{ab}
= &
\int\!\!\dd^d\mathbf{x}\,\partial_t^{2j}\tilde{T}^{(a}\hat{x}^{bL-1)} r^{2j}\,\hat{\partial}_{L-1} E_{ab}\\
& \nm
+\frac{2 \ell}{\ell+1}\,\underset{bL-1}{\mathcal{S}}\left(\int\!\!\dd^d\mathbf{x}\,\partial_t^{2j}\tilde{T}^{[a}\hat{x}^{b]L-1} r^{2j}\right)\hat{\partial}_{L-1} E_{ab}\,.
\end{align}
\end{subequations}
Using the Maxwell-like equations~\eqref{eq:MaxwellGrav}, relations displayed in the appendices, and removing traces with ~\eqref{eqTFone}, one can irreducibly reduce all pieces appearing in those expressions as
\begin{subequations}
\begin{align}
T^{0(a}\hat{x}^{b L-2)}\,\hat{\partial}_{L-2}E_{a b}
= &\, \nm
T^{0\langle a}\hat{x}^{b L-2\rangle}\,\hat{\partial}_{L-2}E_{a b} \\
& \nm
-\frac{4\left(\ell-2 \right)\left(\ell-3 \right)}{\ell\left(d+2\ell-4\right)\left(d+2\ell-8\right)}\,\tilde{T}^0\hat{x}^{a b L-4} \, \hat{\partial}_{L-4}\partial^2_t E_{ab}\\
&
+\frac{4\left(\ell-2 \right)^2\left(\ell-3 \right)r^2}{\ell\left(d+2\ell-4\right)\left(d+2\ell-6\right)\left(d+2\ell-8\right)}\,T^{0\langle a}\hat{x}^{b L-4 \rangle} \, \hat{\partial}_{L-4}\partial^2_t E_{ab}\,,\\
T^{0a}\hat{x}^{bcL-2}\,\hat{\partial}_{L-2}B_{a|bc}
= & \,\nm
\left[T^{0a}\hat{x}^{bcL-2} \right]^{\text{TF}}\,\hat{\partial}_{L-2}B_{a|bc}\\
& \nm
+\frac{\ell-2 }{d+\ell-3 }\,\tilde{T}^{0}\hat{x}^{abL-3}\,\hat{\partial}_{L-3}\partial_t E_{ab}\\
& 
-\frac{\left(\ell-1\right)\left(\ell-2\right)r^2}{\left(d+\ell-3\right)\left(d+2\ell-4\right)}\,T^{0\langle a}\hat{x}^{b L-3\rangle}\hat{\partial}_{L-3}\partial_tE_{ab}\,,\\
\tilde{T}^{(a}\hat{x}^{b L-2)}\,\hat{\partial}_{L-2}E_{a b}
= &\, \nm
\tilde{T}^{\langle a}\hat{x}^{b L-2\rangle}\,\hat{\partial}_{L-2}E_{a b}\\
& \nm
-\frac{4\left(\ell-2 \right)\left(\ell-3 \right)}{\ell\left(d+2\ell-4\right)\left(d+2\ell-8\right)}\,\tilde{T}\hat{x}^{a b L-4} \, \hat{\partial}_{L-4}\partial^2_t E_{ab}\\
&
+\frac{4\left(\ell-2 \right)^2\left(\ell-3 \right)r^2}{\ell\left(d+2\ell-4\right)\left(d+2\ell-6\right)\left(d+2\ell-8\right)}\,\tilde{T}^{\langle a}\hat{x}^{b L-4 \rangle} \, \hat{\partial}_{L-4}\partial^2_t E_{ab}\,,\\
\tilde{T}^{a}\hat{x}^{bcL-2}\,\hat{\partial}_{L-2}B_{a|bc}
= & \, \nm
\left[\tilde{T}^{a}\hat{x}^{bcL-2} \right]^{\text{TF}}\,\hat{\partial}_{L-2}B_{a|bc}\\
& \nm
+\frac{\ell-2}{d+\ell-3}\,\tilde{T}\hat{x}^{abL-3}\,\hat{\partial}_{L-3}\partial_t E_{ab}\\
& 
-\frac{\left(\ell-1\right)\left(\ell-2\right)r^2}{\left(d+\ell-3\right)\left(d+2\ell-4\right)}\,\tilde{T}^{\langle a}\hat{x}^{b L-3\rangle}\,\hat{\partial}_{L-3}\partial_tE_{ab}\,,\\
\underset{b L-2}{\mathcal{S}}\, T^{0[a}\hat{x}^{b]L-2}\,\hat{\partial}_{L-2}E_{ab}
= & \nm
\, -\frac{ \ell-2}{2\left( \ell-1\right)}\,\left[T^{0a}\hat{x}^{bc L-3}\right]^\text{TF}\hat{\partial}_{L-3}\partial_t B_{a|bc}\\
&\nm
-\frac{\left(\ell-2\right)\left(\ell-3\right)\left(\ell-4\right)}{2\left(\ell-1\right)\left(d+\ell-4\right)\left(d+2\ell-8\right)}\, \tilde{T}^0 \hat{x}^{a b L-4}\,\hat{\partial}_{L-4}\partial^2_t E_{a b}\\
&
+\frac{\left(\ell-2\right)^2\left(\ell-3\right)\left(\ell-4\right) r^2}{2\left(\ell-1\right)\left(d+\ell-4\right)\left(d+2\ell-6\right)\left(d+2\ell-8\right)}\,T^{0\langle a}\hat{x}^{b L-4\rangle}\,\hat{\partial}_{L-4}\partial^2_t E_{ab}\,,\\
\underset{b L-2}{\mathcal{S}}\, \tilde{T}^{[a}\hat{x}^{b]L-2}\,\hat{\partial}_{L-2}E_{ab}
= & \nm
-\frac{\ell-2}{2\left( \ell-1\right)}\,\left[\tilde{T}^{a}\hat{x}^{bc L-3}\right]^\text{TF}\hat{\partial}_{L-3}\partial_t B_{a|bc}\\
&\nm
-\frac{\left(\ell-2\right)\left(\ell-3\right)\left(\ell-4\right)}{2\left(\ell-1\right)\left(d+\ell-4\right)\left(d+2\ell-8\right)}\, \tilde{T} \hat{x}^{a b L-4}\,\hat{\partial}_{L-4}\partial^2_t E_{a b}\\
& 
+\frac{\left(\ell-2\right)^2\left(\ell-3\right)\left(\ell-4\right) r^2}{2\left(\ell-1\right)\left(d+\ell-4\right)\left(d+2\ell-6\right)\left(d+2\ell-8\right)}\,\tilde{T}^{\langle a}\hat{x}^{b L-4\rangle}\,\hat{\partial}_{L-4}\partial^2_t E_{ab}\,.
\end{align}
\end{subequations}
After some manipulation, we recover the irreducible expressions of the $T^{0a}$ and $\tilde{T}^a$ sectors, displayed in ~\eqref{ST0a} and~\eqref{STa}, namely
\begin{subequations}
\begin{align}
S^{T^{0a}}_{\text{rad}}
= & \nm \,
2\int\!\!\dd t\,\sum_{\ell=2}^{\infty}\,\sum_{j=0}^{\infty}\,\frac{\Ldlj\,\left(\ell-1\right)}{\blmj{1}\,\left(d+\ell-2\right)}\,\int\!\!\dd^d\mathbf{x}\,\partial_t^{2j+1}T^{0\langle a}\hat{x}^{bL-2\rangle} r^{2j+2}\,\hat{\partial}_{L-2} E_{ab}\\
& \nm 
- 2\int\!\!\dd t\,\sum_{\ell=2}^{\infty}\,\sum_{j=0}^{\infty}\,\frac{\Ldlj\,\left(\ell-1\right)}{\blmj{1}\,\left(d+\ell-2\right)}\,\int\!\!\dd^d\mathbf{x}\,\partial_t^{2j+1}\tilde{T}^{0}\hat{x}^{abL-2} r^{2j}\,\hat{\partial}_{L-2} E_{ab}\\
& 
+ 2\int\!\!\dd t\,\sum_{\ell=2}^{\infty}\,\sum_{j=0}^{\infty}\,\frac{\Ldlj\,\ell\left(\ell-1\right)}{\ell!\,\left(\ell+2j-1\right)\left(\ell+1\right)}\left[\int\!\!\dd^d\mathbf{x}\,\partial_t^{2j}T^{0 a}\hat{x}^{bcL-2} r^{2j}\right]^{\text{TF}}\hat{\partial}_{L-2} B_{a|bc}\,,\\
S^{\tilde{T}^{a}}_{\text{rad}}
= & \nm
-2\int\!\!\dd t\,\sum_{\ell=2}^{\infty}\,\sum_{j=0}^{\infty}\,\frac{\Ldlj}{\glmj{2}}\left(1+\frac{2j}{d+\ell-2}\right)\,\int\!\!\dd^d\mathbf{x}\,\partial_t^{2j}\tilde{T}^{\langle a}\hat{x}^{bL-2\rangle} r^{2j}\,\hat{\partial}_{L-2} E_{ab}\\
& \nm
+ 4\int\!\!\dd t\,\sum_{\ell=2}^{\infty}\,\sum_{j=0}^{\infty}\,\frac{\Ldlj\,j}{\glmj{2}\,\left(d+\ell-2\right)}\,\int\!\!\dd^d\mathbf{x}\,\partial_t^{2j}\tilde{T}\hat{x}^{abL-2} r^{2j-2}\,\hat{\partial}_{L-2} E_{ab}\\
& 
- 2\int\!\!\dd t\,\sum_{\ell=2}^{\infty}\,\sum_{j=0}^{\infty}\,\frac{\Ldlj\,\left(\ell-1\right)}{\glmj{1}\,\left(\ell+1\right)}\left[\int\!\!\dd^d\mathbf{x}\,\partial_t^{2j+1}\tilde{T}^{a}\hat{x}^{bcL-2} r^{2j}\right]^{\text{TF}}\hat{\partial}_{L-2} B_{a|bc}\,,
\end{align}
\end{subequations}

\subsubsection{Tensor sector}\label{sec:app_tensor}
Let us now turn to the $T^{ab}$ sector, displayed in ~\eqref{STab},
\begin{align}\label{STab_app_nonSTF}
S^{T^{ab}}_{\text{rad}}
= &  \nm
\int\!\!\dd t\,\sum_{\ell=2}^{\infty}\,\sum_{j=0}^{\infty}\,\frac{\Ldlj\,\left(d-2\right)}{\glmj{2}}\int\!\!\dd^d\mathbf{x}\,\partial_t^{2j}T^{ab}\hat{x}^{cdL-2} r^{2j}\,\hat{\partial}_{L-2} \mathcal{W}_{acbd}\\
& \nm
- 2\int\!\!\dd t\,\sum_{\ell=1}^{\infty}\,\sum_{j=0}^{\infty}\,\frac{\Ldlj\,\ell\left(d-3\right)}{\glj\,\left(d+2\ell+j\right)}\,\int\!\!\dd^d\mathbf{x}\,\partial_t^{2j+1}T^{ab}\hat{x}^{cL-1} r^{2j+2}\,\hat{\partial}_{L-1} B_{c|ab}\\
& 
+ \int\!\!\dd t\,\sum_{\ell,\,j=0}^{\infty}\,\frac{\Ldlj}{\glj}\left(1+\frac{2j\left(d-3\right)}{d+2\ell+2j}\right)\int\!\!\dd^d\mathbf{x}\,\partial_t^{2j}T^{ab}\hat{x}^L r^{2j+2}\,\hat{\partial}_L E_{ab}\,.
\end{align}
In order to apply the formulas of App.~\ref{sec:app_irred_decomp}, we need to take the STF part of $T^{ab}$ (as the tensor $\mathcal{T}^{abL}$ entering ~\eqref{eqSymmAnti2} and~\eqref{eqTFtwo} has to be separately STF in its two first indices, as well as its $\ell$ other).
Hopefully each of $E_{ab}$, $B_{c\vert ab}$ and $\mathcal{W}_{acbd}$ are traceless.
Hence, one can safely replace $T^{ab}$ by its STF part, $\hat{T}^{ab}$, in ~\eqref{STab_app_nonSTF}
\begin{align}\label{STab_app}
S^{T^{ab}}_{\text{rad}}
= &  \nm
\int\!\!\dd t\,\sum_{\ell=2}^{\infty}\,\sum_{j=0}^{\infty}\,\frac{\Ldlj\,\left(d-2\right)}{\glmj{2}}\int\!\!\dd^d\mathbf{x}\,\partial_t^{2j}\hat{T}^{ab}\hat{x}^{cdL-2} r^{2j}\,\hat{\partial}_{L-2} \mathcal{W}_{acbd}\\
& \nm
- 2\int\!\!\dd t\,\sum_{\ell=1}^{\infty}\,\sum_{j=0}^{\infty}\,\frac{\Ldlj\,\ell\left(d-3\right)}{\glj\,\left(d+2\ell+j\right)}\,\int\!\!\dd^d\mathbf{x}\,\partial_t^{2j+1}\hat{T}^{ab}\hat{x}^{cL-1} r^{2j+2}\,\hat{\partial}_{L-1} B_{c|ab}\\
& 
+ \int\!\!\dd t\,\sum_{\ell,\,j=0}^{\infty}\,\frac{\Ldlj}{\glj}\left(1+\frac{2j\left(d-3\right)}{d+2\ell+2j}\right)\int\!\!\dd^d\mathbf{x}\,\partial_t^{2j}\hat{T}^{ab}\hat{x}^L r^{2j+2}\,\hat{\partial}_L E_{ab}\,.
\end{align}
Once again, the first line is nearly in the sought form, only its trace needs to be removed, which is to be done with the help of ~\eqref{eqTFtwo}
\begin{align}\label{eq:AppDecompTabW}
\hat{T}^{ab}\hat{x}^{cdL}\,\hat{\partial}_L\mathcal{W}_{acbd}
= & \, \nm
\left[T^{ab} \hat{x}^{cdL}\right]^\text{TF}\hat{\partial}_L \mathcal{W}_{acbd} \\
& \nm
+ \frac{2(d-3)\,\ell}{(d-2)(d+\ell)}\,\left[\tilde{T}^c\hat{x}^{abL-1}\right]^\text{TF}\hat{\partial}_{L-1}\partial_tB_{c\vert ab}\\
& \nm
- \frac{2(d-3)\,\ell\,(\ell+1)\,r^2}{(d-2)(d+\ell)(d+2\ell)}\,\left[T^{c\langle a}\hat{x}^{bL-1\rangle}\right]^\text{TF}\hat{\partial}_{L-1}\partial_tB_{c\vert ab}\\
& \nm 
+ \frac{(d-3)\,(\ell-1)\,\ell}{(d-2)(d+\ell-2)(d+\ell-1)}\left(\tilde{T} - \frac{r^2}{d+2\ell}\,T^p_{\, p}\right)\hat{x}^{abL-2}\,\hat{\partial}_{L-2}\partial_t^2E_{ab}\\
& \nm
- \frac{2(d-3)\,(\ell-1)\,\ell^2\,r^2}{(d-2)(d+\ell-2)(d+\ell-1)(d+2\ell)}\,\tilde{T}^{\langle a}\hat{x}^{bL-2\rangle}
\hat{\partial}_{L-2}\partial_t^2E_{ab}\\
& 
+ \frac{(d-3)\,(\ell-1)^2\,\ell^2\,r^4}{(d-2)(d+\ell-2)(d+\ell-1)(d+2\ell-2)(d+2\ell)}\,T^{\langle ab}\hat{x}^{L-2\rangle}\,\hat{\partial}_{L-2}\partial_t^2E_{ab}\,.
\end{align}
As for the second line, after some manipulation, one can irreducibly decompose
\begin{align}\label{eq:AppDecompTabB}
\hat{T}^{ab}\hat{x}^{cL}\,\hat{\partial}_LB_{c|ab}
= & \, \nm
\frac{\ell}{\ell+1}\left[T^{ab} \hat{x}^{cdL-1}\right]^\text{TF}\hat{\partial}_{L-1}\partial_t \mathcal{W}_{acbd}\\
& \nm
- \frac{\ell+2}{\ell+1}\left[T^{c\langle a}\hat{x}^{bL\rangle}\right]^\text{TF}\hat{\partial}_LB_{c\vert ab}\\
& \nm
- \frac{(\ell-1)\ell^2(d+3\ell+3)\,r^2}{(\ell+1)(d+\ell-1)(d+2\ell-2)(d+2\ell)} \left[T^{c\langle a}\hat{x}^{bL-2\rangle}\right]^\text{TF}\hat{\partial}_{L-2}\partial_t^2B_{c\vert ab}\\
& \nm
+ \frac{(\ell-1)\ell(d+3\ell+3)}{(\ell+1)(d+\ell-1)(d+2\ell)}\left[\tilde{T}^c\hat{x}^{abL-2}\right]^\text{TF}\hat{\partial}_{L-2}\partial_t^2B_{c\vert ab}
\\
& \nm 
+ \frac{d\,\ell^2\,r^2}{(d-2)(d+\ell-1)(d+2\ell-2)}\,T^{\langle ab}\hat{x}^{L-1\rangle}\hat{\partial}_{L-1}\partial_tE_{ab}\\
& \nm
+ \frac{(\ell-2)^2(\ell-1)^2\ell\,r^4}{(d+\ell-3)(d+\ell-2)(d+2\ell-2)^2(d+2\ell-4)}\,T^{\langle ab}\hat{x}^{L-3\rangle}\hat{\partial}_{L-3}\partial_t^3E_{ab}\\
& \nm 
- \frac{d\,\ell}{(d-2)(d+\ell-1)}\,\tilde{T}^{\langle a}\hat{x}^{bL-1\rangle}\hat{\partial}_{L-1}\partial_tE_{ab}
\\
& \nm 
- \frac{2(\ell-2)(\ell-1)^2\ell\,r^2}{(d+\ell-3)(d+\ell-2)(d+2\ell-2)^2}\,\tilde{T}^{\langle a}\hat{x}^{bL-3\rangle}\hat{\partial}_{L-3}\partial_t^3E_{ab}
\\
& \nm 
+ \frac{\ell}{(d-2)(d+\ell-1)}\,T^p_{\, p}\,\hat{x}^{abL-1}\hat{\partial}_{L-1}\partial_tE_{ab}\\
& \nm 
- \frac{(\ell-2)(\ell-1)\ell\,r^2}{(d+\ell-3)(d+\ell-2)(d+2\ell-2)^2}\,T^p_{\, p}\,\hat{x}^{abL-3}\hat{\partial}_{L-3}\partial_t^3E_{ab}\\
& 
+ \frac{(\ell-2)(\ell-1)\ell}{(d+\ell-3)(d+\ell-2)(d+2\ell-2)}\,\tilde{T}\,\hat{x}^{abL-3}\hat{\partial}_{L-3}\partial_t^3E_{ab}\,.
\end{align}
In order to apply this formula to ~\eqref{STab_app}, one simply need to downgrade the value of $\ell$ by one.
Finally, for the last line, let us first (anti-)symmetrize it by using ~\eqref{eqSymmAnti2}
\begin{align}
\int\!\!\dd^d\mathbf{x}\,\partial_t^{2j}\hat{T}^{ab}\hat{x}^L r^{2j+2}\,\hat{\partial}_L E_{ab}
 &= \nm
\int\!\!\dd^d\mathbf{x}\,\partial_t^{2j}\hat{T}^{(ab}\hat{x}^{L)} r^{2j+2}\,\hat{\partial}_L E_{ab}\\
& \nm
+ \frac{2\ell \left( \ell+3\right)}{\left( \ell+1\right)\left( \ell+2\right)}\int\!\!\dd^d\mathbf{x}\,\partial_t^{2j}\hat{T}^{ab}\hat{x}^{L} r^{2j+2}\,\hat{\partial}_{L-1[i_{\ell}} E_{a]b}\\
& 
+ \frac{2\ell \left( \ell-1\right)}{\left( \ell+1\right)\left( \ell+2\right)}\int\!\!\dd^d\mathbf{x}\,\partial_t^{2j}\hat{T}^{a i_{\ell-1}}\hat{x}^{b i_{\ell}L-2} r^{2j+2}\,\hat{\partial}_{L-1[i_{\ell}} E_{a]b}\,.
\end{align}
Working out those coefficients with the set of relations at hand, we find
\begin{subequations}
\begin{align}
\hat{T}^{(a b}\hat{x}^{L)}\,\hat{\partial}_L E_{a b}
= & \, \nm
T^{\langle a b}\hat{x}^{L\rangle}\,\hat{\partial}_LE_{a b}\\
&\nm
+\frac{8 \ell^2 \left(\ell-1\right)^2 r^2}{\left(\ell+1 \right)\left(\ell+2 \right)\left(d+2\ell \right)\left(d+2\ell -4\right)^2}\, T^{\langle ab}\hat{x}^{L-2\rangle}\,\hat{\partial}_{L-2}\partial^2_t E_{ab}\\
&\nm
+\frac{8 \ell \left(\ell-1\right)\left(\ell-2\right)^2\left(\ell-3\right)^2 r^4}{\left(\ell+1 \right)\left(\ell+2 \right)\left(d+2\ell-2 \right)\left(d+2\ell -4\right)^3\left(d+2\ell -6\right)^2}\, T^{\langle ab}\hat{x}^{L-4\rangle}\,\hat{\partial}_{L-4}\partial^4_t E_{ab}\\
&\nm
-\frac{8 \ell^2 \left(\ell-1\right)}{\left(\ell+1 \right)\left(\ell+2 \right)\left(d+2\ell \right)\left(d+2\ell -4\right)}\, \tilde{T}^{\langle a}\hat{x}^{b L-2\rangle}\,\hat{\partial}_{L-2}\partial^2_t E_{ab}\\
&\nm
-\frac{16 \ell \left(\ell-1\right)\left(\ell-2\right)^2\left(\ell-3\right) r^2}{\left(\ell+1 \right)\left(\ell+2 \right)\left(d+2\ell-2 \right)\left(d+2\ell -4\right)^3\left(d+2\ell -6\right)}\, \tilde{T}^{\langle a}\hat{x}^{b L-4\rangle}\,\hat{\partial}_{L-4}\partial^4_t E_{ab}\\
& \nm
+\frac{8 \ell^2 \left(\ell-1\right)}{\left(\ell+1 \right)\left(\ell+2 \right)d\left(d+2\ell \right)\left(d+2\ell -4\right)}\, T^{aa}\,\hat{x}^{ab L-2}\,\hat{\partial}_{L-2}\partial^2_t E_{ab}\\
& \nm
-\frac{8 \ell \left(\ell-1\right)\left(\ell-2\right)\left(\ell-3\right) r^2}{\left(\ell+1 \right)\left(\ell+2 \right)\left(d+2\ell-2 \right)\left(d+2\ell -4\right)^3\left(d+2\ell -6\right)}\,T^{aa}\,\hat{x}^{ab L-4}\,\hat{\partial}_{L-4}\partial^4_t E_{ab}\\
& 
+\frac{8 \ell \left(\ell-1\right)\left(\ell-2\right)\left(\ell-3\right)}{\left(\ell+1 \right)\left(\ell+2 \right)\left(d+2\ell-2 \right)\left(d+2\ell -4\right)^2\left(d+2\ell -6\right)}\,\tilde{T}\,\hat{x}^{ab L-4}\,\hat{\partial}_{L-4}\partial^4_t E_{ab}\,,\\
\hat{T}^{ab}\hat{x}^L\,\hat{\partial}_{L-1[i_{\ell}}E_{a]b}
= &\, \nm
\frac{\ell-1}{2\,\ell}\left[T^{ab} \hat{x}^{cdL-2}\right]^\text{TF}\hat{\partial}_{L-2}\partial_t^2 \mathcal{W}_{acbd}\\
& \nm
- \frac{\ell+1}{2\,\ell}\left[T^{c\langle a}\hat{x}^{bL-1\rangle}\right]^\text{TF}\hat{\partial}_{L-1}\partial_tB_{c\vert ab}
\\
& \nm
- \frac{(\ell-2)(\ell-1)^2[2(\ell^2-1)+(d-3)\ell]\,r^2}{\ell\,(d+\ell-2)(d+2\ell-2)(d+2\ell-4)^2} \left[T^{c\langle a}\hat{x}^{bL-3\rangle}\right]^\text{TF}\hat{\partial}_{L-3}\partial_t^3B_{c\vert ab}
\\
& \nm
+ \frac{(\ell-2)(\ell-1)[2(\ell^2-1)+(d-3)\ell]}{\ell\,(d+\ell-2)(d+2\ell-2)(d+2\ell-4)}\left[\tilde{T}^c\hat{x}^{abL-3}\right]^\text{TF}\hat{\partial}_{L-3}\partial_t^3B_{c\vert ab}
\\
& \nm
+ \frac{(\ell-1)^2(d\ell+2\ell-4)\,r^2}{2(d-2)(d+\ell-2)(d+2\ell-4)^2}\,T^{\langle ab}\hat{x}^{L-2\rangle}\hat{\partial}_{L-2}\partial_t^2E_{ab}
\\
& \nm
+ \frac{(\ell-3)^2(\ell-2)^2(\ell-1)^2\,r^4}{2(d+\ell-3)(d+\ell-4)(d+2\ell-4)^3(d+2\ell-6)}\,T^{\langle ab}\hat{x}^{L-4\rangle}\hat{\partial}_{L-4}\partial_t^4E_{ab}
\\
& \nm
- \frac{(\ell-1)(d\ell+2\ell-4)}{2(d-2)(d+\ell-2)(d+2\ell-4)}\,\tilde{T}^{\langle a}\hat{x}^{bL-2\rangle}\hat{\partial}_{L-2}\partial_t^2E_{ab}
\\
& \nm
- \frac{(\ell-3)(\ell-2)^2(\ell-1)^2\,r^2}{(d+\ell-3)(d+\ell-4)(d+2\ell-4)^3}\,\tilde{T}^{\langle a}\hat{x}^{bL-4\rangle}\hat{\partial}_{L-4}\partial_t^4E_{ab}
\\
& \nm
+ \frac{(\ell-1)(d\ell+2\ell-4)}{2d\,(d-2)(d+\ell-2)(d+2\ell-4)}\,T^p_{\, p}\,\hat{x}^{abL-2}\hat{\partial}_{L-2}\partial_t^2E_{ab}
\\
& \nm
- \frac{(\ell-3)(\ell-2)(\ell-1)^2\,r^2}{2(d+\ell-3)(d+\ell-4)(d+2\ell-4)^3}\,T^p_{\, p}\hat{x}^{abL-4}\hat{\partial}_{L-4}\partial_t^4E_{ab}
\\
& 
+ \frac{(\ell-3)(\ell-2)(\ell-1)^2}{2(d+\ell-3)(d+\ell-4)(d+2\ell-4)^2}\,\tilde{T}\hat{x}^{abL-4}\hat{\partial}_{L-4}\partial_t^4E_{ab}\,,
\\
\hat{T}^{ai_{\ell-1}}\hat{x}^{bi_\ell L-2}\hat{\partial}_{L-1[i_\ell}E_{a]b}
 &= \nm
- \frac{1}{2\,\ell}\left[T^{ab} \hat{x}^{cdL-2}\right]^\text{TF}\hat{\partial}_{L-2}\partial_t^2 \mathcal{W}_{acbd}
\\
& \nm
- \frac{\ell+1}{2\,\ell}\left[T^{c\langle a}\hat{x}^{bL-1\rangle}\right]^\text{TF}\hat{\partial}_{L-1}\partial_tB_{c\vert ab}
\\
& \nm
- \frac{(\ell-2)(\ell-1)(2d-\ell^2+3\ell-2)\, r^2}{\ell(d+\ell-2)(d+2\ell-2)(d+2\ell-4)^2}\left[T^{c\langle a}\hat{x}^{bL-3\rangle}\right]^\text{TF}\hat{\partial}_{L-3}\partial_t^3B_{c\vert ab}
\\
& \nm
+ \frac{(\ell-2)(2d-\ell^2+3\ell-2)}{\ell(d+\ell-2)(d+2\ell-2)(d+2\ell-4)}\left[\tilde{T}^c\hat{x}^{abL-3}\right]^\text{TF}\hat{\partial}_{L-3}\partial_t^3B_{c\vert ab}
\\
& \nm
+ \frac{(\ell-1)^2[(d-2)(\ell-2)-2d]\,r^2}{2(d-2)(d+\ell-2)(d+2\ell-4)^2}\,T^{\langle ab}\hat{x}^{L-2\rangle}\hat{\partial}_{L-2}\partial_t^2E_{ab}
\\
& \nm
+ \frac{(d-2)(\ell-2)^2(\ell-3)^3\,r^4}{2(d+\ell-3)(d+\ell-4)(d+2\ell-4)^3(d+2\ell-6)^2}\,T^{\langle ab}\hat{x}^{L-4\rangle}\hat{\partial}_{L-4}\partial_t^4E_{ab}
\\
& \nm
- \frac{(\ell-1)[(d-2)(\ell-2)-2d]}{2(d-2)(d+\ell-2)(d+2\ell-4)}\,\tilde{T}^{\langle a}\hat{x}^{bL-2\rangle}\hat{\partial}_{L-2}\partial_t^2E_{ab}
\\
& \nm
- \frac{(d-2)(\ell-2)^2(\ell-3)^2\,r^2}{(d+\ell-3)(d+\ell-4)(d+2\ell-4)^3(d+2\ell-6)}\,\tilde{T}^{\langle a}\hat{x}^{bL-4\rangle}\hat{\partial}_{L-4}\partial_t^4E_{ab}
\\
& \nm
+ \frac{(\ell-1)[(d-2)(\ell-2)-2d]}{2d(d-2)(d+\ell-2)(d+2\ell-4)}\,T^p_{\, p}\hat{x}^{abL-2}\hat{\partial}_{L-2}\partial_t^2E_{ab}
\\
& \nm
- \frac{(d-2)(\ell-2)(\ell-3)^2\,r^2}{2(d+\ell-3)(d+\ell-4)(d+2\ell-4)^3(d+2\ell-6)}\,T^p_{\, p}\hat{x}^{abL-4}\hat{\partial}_{L-4}\partial_t^4E_{ab}
\\ 
& 
+ \frac{(d-2)(\ell-2)(\ell-3)^2\,r^4}{2(d+\ell-3)(d+\ell-4)(d+2\ell-4)^2(d+2\ell-6)}\,\tilde{T}\hat{x}^{abL-4}\hat{\partial}_{L-4}\partial_t^4E_{ab}\,.
\end{align}
\end{subequations}
Injecting all those relations into ~\eqref{STab_app}, we recover the result displayed in ~\eqref{STabfin}, that we recall here
\begin{align}
S^{T^{ab}}_{\text{rad}}
= & \nm
\int\!\!\dd t\,\sum_{\ell=2}^{\infty}\,\sum_{j=0}^{\infty}\,\frac{\Ldlj}{\glmj{2}}\int\!\!\dd^d\mathbf{x}\,\partial_t^{2j}T^{\langle ab}\hat{x}^{L-2\rangle} r^{2j+2}\,\hat{\partial}_{L-2} E_{ab}\\
& \nm
+ 2 \int\!\!\dd t\,\sum_{\ell=2}^{\infty}\,\sum_{j=0}^{\infty}\,\frac{\Ldlj\,j\left(d-1\right)\left(d+2\ell+2j-1\right)}{\glmj{2}\,\left(d+\ell-1\right)\left(d+\ell-2\right)}\int\!\!\dd^d\mathbf{x}\,\partial_t^{2j}T^{\langle ab}\hat{x}^{L-2\rangle} r^{2j+2}\,\hat{\partial}_{L-2} E_{ab}\\
& \nm
+4 \int\!\!\dd t\,\sum_{\ell=2}^{\infty}\,\sum_{j=0}^{\infty}\,\frac{\Ldlj\,j}{\glmj{2}\left(d+\ell-2\right)}\left(1-\frac{\left(d-1\right)\left(d+2\ell+2j-1\right)}{\left(d+\ell-1\right)}\right)\int\!\!\dd^d\mathbf{x}\,\partial_t^{2j}T^{\langle a}\hat{x}^{b L-2\rangle} r^{2j}\,\hat{\partial}_{L-2} E_{ab}\\
& \nm
+2 \int\!\!\dd t\,\sum_{\ell=2}^{\infty}\,\sum_{j=0}^{\infty}\,\frac{\Ldlj\,j \left(d+2\ell+2j-1\right)}{\glmj{2} \,\left(d+\ell-1\right)\left(d+\ell-2\right)}\,\int\!\!\dd^d\mathbf{x}\,\partial_t^{2j}T^p_{\, p}\hat{x}^{ab L-2} r^{2j}\,\hat{\partial}_{L-2} E_{ab}\\
& \nm
+ 4\int\!\!\dd t\,\sum_{\ell=2}^{\infty}\,\sum_{j=0}^{\infty}\,\frac{\Ldlj\,j\left[2\left(j-1\right)\left(d-2\right)+\left(d+2\ell\right)\left(d-3\right)\right]}{\glmj{2}\,\left(d+\ell-1 \right)\left(d+\ell-2 \right)}\,\int\!\!\dd^d\mathbf{x}\,\partial_t^{2j}\tilde{T}\hat{x}^{ab L-2} r^{2j-2}\,\hat{\partial}_{L-2} E_{ab}\\
& \nm
+ 2\int\!\!\dd t\,\sum_{\ell=2}^{\infty}\,\sum_{j=0}^{\infty}\,\frac{\Ldlj\,\ell\left( \ell+2j+2\right)\left( d-2\right)}{\glmj{1} \,\left(\ell+1\right)\left(d+\ell-1\right)}\left[\int\!\!\dd^d\mathbf{x}\,\partial_t^{2j+1}T^{a \langle b}\hat{x}^{cL-2\rangle} r^{2j+2}\right]^{\text{TF}}\hat{\partial}_{L-2} B_{a|bc}\\
& \nm
- 2\int\!\!\dd t\,\sum_{\ell=2}^{\infty}\,\sum_{j=0}^{\infty}\,\frac{\Ldlj\,\left(\ell-1\right)\left[2j\left(d-2\right)+\left(\ell+1\right)\left(d-3\right)\right]}{\glmj{1}\,\left(\ell+1\right)\left(d+\ell-1\right)}\left[\int\!\!\dd^d\mathbf{x}\,\partial_t^{2j+1}\tilde{T}^{a}\hat{x}^{b c L-2}r^{2j}\right]^{\text{TF}}\hat{\partial}_{L-2} B_{a|bc}\\
& 
+ \int\!\!\dd t\,\sum_{\ell=2}^{\infty}\,\sum_{j=0}^{\infty}\,\frac{\Ldlj\,\left(\ell+2j\right)\left(\ell+2j+1\right)\left( d-2\right)}{\glmj{2}\,\left(\ell+1 \right)\left(\ell +2\right)}\left[\int\!\!\dd^d\mathbf{x}\,\partial_t^{2j}T^{ab}\hat{x}^{cdL-2} r^{2j}\right]^{\text{TF}}\hat{\partial}_{L-2} \mathcal{W}_{acbd}\,.
\end{align}

\subsubsection{Summing all sectors}\label{sec:app_sum}

The full radiative action is
\begin{equation}
S_\text{rad} = S^{T^{00}}_{\text{rad}}+S^{T^{aa}}_{\text{rad}}+S^{T^{0a}}_{\text{rad}}+S^{\tilde{T}^a}_{\text{rad}}+S^{T^{ab}}_{\text{rad}}\,,
\end{equation}
where the irreducible decompositions of the five terms are displayed in respectively in ~\eqref{ST00},~\eqref{STpp},~\eqref{ST0a},~\eqref{STa} and~\eqref{STabfin}.
Putting everything together, it comes
\begin{align} \label{finalLG}
S_{\text{rad}}
= & \nm
\int\!\!\dd t\,\sum_{\ell=2}^{\infty}\,\sum_{j=0}^{\infty}\,\frac{\Ldlj\,\ell\left(\ell-1\right)}{\ell!\left(\ell+2j-1\right)\left(\ell+2j\right)}\,\int\!\!\dd^d\mathbf{x}\,\partial_t^{2j}T^{00}\hat{x}^{abL-2} r^{2j}\,\hat{\partial}_{L-2} E_{ab}\\
& \nm
- 2\int\!\!\dd t\,\sum_{\ell=2}^{\infty}\,\sum_{j=0}^{\infty}\,\frac{\Ldlj\,\ell\left(\ell-1\right)}{\ell!\left(\ell+2j\right)\left(\ell+2j+2\right)\left(d+\ell-2\right)}\,\int\!\!\dd^d\mathbf{x}\,\partial_t^{2j+1}\tilde{T}^{0}\hat{x}^{abL-2} r^{2j}\,\hat{\partial}_{L-2} E_{ab}\\
& \nm
+ 2\int\!\!\dd t\,\sum_{\ell=2}^{\infty}\,\sum_{j=0}^{\infty}\,\frac{\Ldlj\,\ell\left(\ell-1\right)}{\ell!\left(\ell+2j\right)\left(\ell+2j+2\right)\left(d+\ell-2\right)}\,\int\!\!\dd^d\mathbf{x}\,\partial_t^{2j+1}T^{0\langle a}\hat{x}^{bL-2 \rangle} r^{2j+2}\,\hat{\partial}_{L-2} E_{ab}\\
& \nm
+ 2\int\!\!\dd t\,\sum_{\ell=2}^{\infty}\,\sum_{j=0}^{\infty}\,\frac{\Ldlj\,\ell\left(\ell-1\right)j\left(d+2\ell+2j-1\right)}{\ell!\left(\ell+2j\right)\left(\ell+2j+1\right)\left(d+\ell-1 \right)\left(d+\ell-2\right)}\\
& \nm \hspace{9cm} \times
\int\!\!\dd^d\mathbf{x}\,\partial_t^{2j}\tilde{T}\hat{x}^{abL-2} r^{2j-2}\,\hat{\partial}_{L-2} E_{ab}\\
& \nm
+ \int\!\!\dd t\,\sum_{\ell=2}^{\infty}\,\sum_{j=0}^{\infty}\,\frac{\Ldlj\,\ell\left(\ell-1\right)}{\ell!\left(\ell+2j\right)\left(\ell+2j+1\right)\left(d-2\right)}\,\left(1+\frac{2j\left(d+2\ell+2j-1\right)}{\left(d+\ell-1\right)\left(d+\ell-2\right)}\right)\\
& \nm \hspace{9cm} \times
\int\!\!\dd^d\mathbf{x}\,\partial_t^{2j}T^{aa}\,\hat{x}^{abL-2} r^{2j}\,\hat{\partial}_{L-2} E_{ab}\\
& \nm
- 2\int\!\!\dd t\,\sum_{\ell=2}^{\infty}\,\sum_{j=0}^{\infty}\,\frac{\Ldlj\,\ell\left(\ell-1\right)}{\ell!\left(\ell+2j\right)\left(\ell+2j+1\right)\left(d-2\right)}\,\left(1+\frac{2j\left(d-1\right)\left(d+2\ell+2j-1\right)}{\left(d+\ell-1\right)\left(d+\ell-2\right)}\right)\\
& \nm \hspace{9cm} \times
\int\!\!\dd^d\mathbf{x}\,\partial_t^{2j}\tilde{T}^{\langle a}\hat{x}^{bL-2 \rangle} r^{2j}\,\hat{\partial}_{L-2} E_{ab}\\
& \nm
+ \int\!\!\dd t\,\sum_{\ell=2}^{\infty}\,\sum_{j=0}^{\infty}\,\frac{\Ldlj\,\ell\left(\ell-1\right)}{\ell!\left(\ell+2j\right)\left(\ell+2j+1\right)\left(d-2\right)}\,\left(1+\frac{2j\left(d-1\right)\left(d+2\ell+2j-1\right)}{\left(d+\ell-1\right)\left(d+\ell-2\right)}\right)\\
& \nm \hspace{9cm} \times
\int\!\!\dd^d\mathbf{x}\,\partial_t^{2j}\tilde{T}^{\langle a b}\hat{x}^{L-2 \rangle} r^{2j+2}\,\hat{\partial}_{L-2} E_{ab}\\
& \nm
+ 2\int\!\!\dd t\,\sum_{\ell=2}^{\infty}\,\sum_{j=0}^{\infty}\,\frac{\Ldlj\,\ell\left(\ell-1\right)}{\left(\ell+1\right)!\left(\ell+2j-1\right)}\,\left[\int\!\!\dd^d\mathbf{x}\,\partial_t^{2j}T^{0a}\hat{x}^{bcL-2} r^{2j}\right]^{\text{TF}}\hat{\partial}_{L-2} B_{a|bc}\\
& \nm
- 2\int\!\!\dd t\,\sum_{\ell=2}^{\infty}\,\sum_{j=0}^{\infty}\,\frac{\Ldlj\,\ell\left(\ell-1\right)}{\left(\ell+1\right)!\left(\ell+2j+1\right)\left( d+\ell-1\right)}\,\left[\int\!\!\dd^d\mathbf{x}\,\partial_t^{2j+1}\tilde{T}^{a}\hat{x}^{bcL-2} r^{2j}\right]^{\text{TF}}\hat{\partial}_{L-2} B_{a|bc}\\
& \nm
+ 2\int\!\!\dd t\,\sum_{\ell=2}^{\infty}\,\sum_{j=0}^{\infty}\,\frac{\Ldlj\,\ell^2}{\left(\ell+1\right)!\left(\ell+2j+1\right)\left(d+\ell-1 \right)}\,\left[\int\!\!\dd^d\mathbf{x}\,\partial_t^{2j+1}T^{a\langle b}\hat{x}^{cL-2\rangle} r^{2j+2}\right]^{\text{TF}}\hat{\partial}_{L-2} B_{a|bc}\\
& 
+ \int\!\!\dd t\,\sum_{\ell=2}^{\infty}\,\sum_{j=0}^{\infty}\,\frac{\Ldlj\,\left(\ell-1\right)}{\left(\ell+1\right)!}\,\left[\int\!\!\dd^d\mathbf{x}\,\partial_t^{2j}T^{a b}\hat{x}^{cdL-2} r^{2j}\right]^{\text{TF}}\hat{\partial}_{L-2} \mathcal{W}_{acbd}\,.
\end{align}
Implementing the conservation laws~\eqref{Tcons} to replace the coefficients involving $T^{i_{\ell}i_{\ell-1}}$, $\tilde{T}^{i_{\ell}}$, $T^{0 i_{\ell}}$ and $T^{a i_{\ell}}$, it finally comes
\begin{align}
S_{\text{rad}}
= & \nm
\int\!\!\dd t\,\sum_{\ell=2}^{\infty}\,\sum_{j=0}^{\infty}\,\frac{\Ldlj}{\ell!}\left(1+\frac{4j\left(d-1\right)\left(d+\ell+j-2\right)}{\left(d-2\right)\left(d+\ell-1\right)\left(d+\ell-2\right)}\right)\int\!\!\dd^d\mathbf{x}\,\partial_t^{2j}T^{00}\hat{x}^{abL-2} r^{2j}\,\hat{\partial}_{L-2} E_{ab}\\
& \nm
-2\int\!\!\dd t\,\sum_{\ell=2}^{\infty}\,\sum_{j=0}^{\infty}\,\frac{\Ldlj\,\left(d-1\right)\left(d+\ell+2j-1\right)}{\ell!\,\left(d-2\right)\left(d+\ell-1\right)\left(d+\ell-2\right)}\int\!\!\dd^d\mathbf{x}\,\partial_t^{2j+1}\tilde{T}^{0}\hat{x}^{abL-2} r^{2j}\,\hat{\partial}_{L-2} E_{ab}\\
& \nm
+\int\!\!\dd t\,\sum_{\ell=2}^{\infty}\,\sum_{j=0}^{\infty}\,\frac{\Ldlj}{\ell!\,\left(d-2\right)}\left(1+\frac{2j\left(d-1\right)}{\left(d+\ell-1\right)\left(d+\ell-2\right)}\right)\int\!\!\dd^d\mathbf{x}\,\partial_t^{2j}T^{aa}\,\hat{x}^{abL-2} r^{2j}\,\hat{\partial}_{L-2} E_{ab}\\
& \nm
+\int\!\!\dd t\,\sum_{\ell=2}^{\infty}\,\sum_{j=0}^{\infty}\,\frac{\Ldlj\,\left(d-1\right)}{\ell!\,\left(d-2\right)\left(d+\ell-1\right)\left(d+\ell-2\right)}\int\!\!\dd^d\mathbf{x}\,\partial_t^{2j+2}\tilde{T}\hat{x}^{abL-2} r^{2j}\,\hat{\partial}_{L-2} E_{ab}\\
& \nm
-2\int\!\!\dd t\,\sum_{\ell=2}^{\infty}\,\sum_{j=0}^{\infty}\,\frac{\Ldlj\,\ell}{\left(\ell+1\right)!\,\left( d+\ell-1\right)}\,\left[\int\!\!\dd^d\mathbf{x}\,\partial_t^{2j+1}\tilde{T}^{a}\hat{x}^{bcL-2} r^{2j}\right]^{\text{TF}}\hat{\partial}_{L-2} B_{a|bc}\\
& \nm
+2\int\!\!\dd t\,\sum_{\ell=1}^{\infty}\,\sum_{j=0}^{\infty}\,\frac{\Ldlj\,\ell}{\left(\ell+1\right)!}\,\ell\left(1+\frac{2j}{d+\ell-1}\right)\left[\int\!\!\dd^d\mathbf{x}\,\partial_t^{2j}T^{0a}\hat{x}^{bcL-2} r^{2j}\right]^{\text{TF}}\hat{\partial}_{L-2} B_{a|bc}\\
&
+ \int\!\!\dd t\,\sum_{\ell=2}^{\infty}\,\sum_{j=0}^{\infty}\,\frac{\Ldlj\,\left(\ell-1\right)}{\left(\ell+1\right)!}\,\left[\int\!\!\dd^d\mathbf{x}\,\partial_t^{2j}T^{a b}\hat{x}^{cdL-2} r^{2j}\right]^{\text{TF}}\hat{\partial}_{L-2} \mathcal{W}_{acbd}\,,
\end{align}
from which we extract our final result, ~\eqref{sourcemultexpGrav} and~\eqref{eq:momentsGrav}.

\bibliography{Ref_ddim.bib}

\end{document}